\documentclass[11pt,USenglish]{article}
\pdfoutput=1

\usepackage[utf8]{inputenc}
\usepackage{geometry}
\usepackage{cancel}
\usepackage[USenglish]{babel}
\usepackage{verbatim}
\usepackage{amsmath,amssymb,amscd,xcolor}
\usepackage{amsfonts}
\usepackage{mathrsfs}
\usepackage{mathtools}
\usepackage{float}
\usepackage{graphicx}
\usepackage{url}
\usepackage{cite}
\usepackage{physics}
\usepackage{braket}
\usepackage{verbatim}
\usepackage[font=footnotesize,labelfont=bf,width=.9\textwidth]{caption}
\usepackage{multirow}
\usepackage{boldline}
\usepackage{enumitem}
\usepackage{wrapfig}
\usepackage{graphicx} 
\usepackage{setspace}
\usepackage{slashed}
\usepackage{dsfont}

\usepackage{color}
\definecolor{darkgreen}{rgb}{0,0.5,0}
\definecolor{darkblue}{rgb}{0,0,0.6}
\definecolor{purple}{rgb}{0.4,.2,0.7}
\definecolor{orange}{rgb}{0.95, 0.5, 0.3}

\usepackage[colorlinks=true,citecolor=darkgreen,linkcolor=darkblue,urlcolor=blue]{hyperref}
\usepackage{cleveref}

\newcommand{\beq}{\begin{equation}}
\newcommand{\eeq}{\end{equation}}
\newcommand{\pa}{\partial}
\newcommand{\mc}[1]{\mathcal{#1}}

\newcommand{\haLP}[1]{P_{i#1-\frac12}^{\lambda-\frac12}}

\newcommand{\slalg}{\mathfrak{sl}(2,\mathbb R)}
\newcommand{\vdm}{\boldsymbol{\Delta}}

\numberwithin{equation}{section}
\numberwithin{table}{section}

\usepackage[normalem]{ulem}

\begin{document}

\begin{spacing}{1.1}

~
\vskip5mm

~
\vskip5mm

\begin{center} {\Huge \textsc \bf Medicine show: A Calogero model with principal series states}

\vskip10mm

Tarek Anous,$^{1}$ Jackson R. Fliss,$^{2}$ and Jeremy van der Heijden$^{3}$
\vskip1em
{\it 1) School of Mathematical Sciences, Queen Mary University of London, Mile End Road, London E1
4NS, United Kingdom} \\
\vskip5mm
{\it 2) Department of Applied Mathematics and Theoretical Physics,} \\
{\it University of Cambridge, Cambridge CB3 0WA, UK}  \\
\vskip5mm
{\it 3) Department of Physics and Astronomy, University of British Columbia, 6224 Agricultural Road, Vancouver, B.C. V6T 1Z1, Canada} \\
\vskip5mm

\tt{ t.anous@qmul.ac.uk, jf768@cam.ac.uk, jeremy.vanderheijden@ubc.ca}

\end{center}

\vskip10mm

\begin{abstract}
The \textit{Calogero model} is an interacting, $N$-particle, $\slalg$-invariant quantum mechanics, whose Hilbert space is furnished by a tower of discrete series modules. The system enjoys both classical and quantum integrability at any $N$ and at any value of the coupling; this is guaranteed by the existence of $N$ mutually-commuting currents, one of them being the Hamiltonian.  In this paper, we alter the Calogero model so that it may accommodate states in the unitary principal series irreducible representation of $\slalg$. Doing so requires changing the domain of the quantum operators---a procedure which succeeds in preserving unitarity and $\slalg$-invariance, but alters the integrability properties of the theory. We explicitly solve the deformed model for $N=2,3$ and outline a procedure for solving the model at general $N$. We expect this deformed model to provide us with general lessons that carry over to other systems with states in the principal series, for example, interacting massive quantum field theories on de Sitter space.
\end{abstract}

\pagebreak

\pagestyle{plain}

\setcounter{tocdepth}{3}
{}
\vfill
\tableofcontents

\section{Introduction}

Interacting quantum field theory (QFT) on a rigid de Sitter background remains a notoriously difficult subject. Being a time-dependent (accelerating) cosmology, de Sitter spacetime has no conserved energy, a fact with drastic implications. As a simple illustration, let us take the notion that all effective theories in flat space come equipped with a cutoff scale $\Lambda_{\rm UV}$ beyond which we are completely ignorant of the particle content, states, etc. Starting from a UV Lagrangian, the procedure that generates the effective theory is to chop the Euclidean path integral up into energy shells, and integrate-out all the shells above $\Lambda_{\rm UV}$. What results is a self-contained theory, so long as we restrict ourselves to only consider processes well below the cutoff scale. 

Following a similar process for generating an effective field theory in de Sitter spacetime runs into both conceptual and technical difficulties related to the fact that Euclidean de Sitter space is compact, and hence its momentum shells are discrete. And indeed it is difficult to imagine what the effective QFT Lagrangian, with an intrinsic cutoff scale $\Lambda_{\rm UV}$ produced via a discrete sum, would entail: physical processes are not constrained by energy conservation, and starting from a generic `light' state, we eventually populate states of arbitrarily-high `mass'\footnote{As specified by the Lagrangian coupling, say.} so long as we allow ourselves to wait long enough \cite{nachtmann1968dynamische, Bros:2010rku,Epstein:2012zz}. A potentially better notion of a cutoff in de Sitter might be a time-band within which we can reliably trust the physical description, rather than an energy scale. This certainly evokes the problem of secular growths (see \cite{Tsamis:1992sx,Tsamis:1996qq,Weinberg:2005vy,Weinberg:2006ac,Burgess:2010dd,Polyakov:2007mm,Burgess:2009bs,Seery:2010kh,Anninos:2014lwa,Senatore:2009cf,Freese:1987vx,Youssef:2013by,Starobinsky:1994bd,Ford:1984hs} for an incomplete list of references discussing infrared issues in de Sitter spacetimes), 
which we will not delve into here. However, this proposed notion of a cutoff is at odds with certain efforts to formulate theories of cosmology via the `cosmological bootstrap' \cite{Maldacena:2011nz,Mata:2012bx,Bzowski:2012ih,Kundu:2014gxa,Arkani-Hamed:2015bza,Arkani-Hamed:2018kmz,Sleight:2019hfp,Hogervorst:2021uvp,DiPietro:2021sjt,Baumann:2022jpr}, in which the correlation functions of interest are anchored at the infinite asymptotic future of the spacetime, far beyond any time-cutoff we may want to impose. 

Solveable models upon which we can build intuition are greatly needed in this context. However, absent a slew of exactly solvable interacting QFTs on de Sitter backgrounds (see \cite{DiPietro:2023inn,Anninos:2024fty}
 for recent progress in this area), our next-best hope is to write down, and hopefully solve, models which share certain features of interest with de Sitter QFT. This is the goal of our paper, and given the quagmire in which this problem exists, we shall approach it from the safe shores of symmetry. 

Much like flat space, $d$-dimensional de Sitter is a maximally symmetric spacetime, albeit one with an intrinsic scale, the de Sitter length $L_{\rm dS}$. This scale allows us to distinguish between heavy $\left(mL_{\rm dS}  > \tfrac{d-1}{2} \right)$ and light $\left(mL_{\rm dS}  < \tfrac{d-1}{2} \right)$ fields and each subcase falls under a \emph{different} unitary irreducible representations (UIR) of the de Sitter algebra $\mathfrak{so}(1,d)$. In what follows we will be interested in the \emph{principal series} representation, describing heavy fields. To simplify our lives, we will consider the simplest case $\mathfrak{so}(1,2)$, the isometry group of two-dimensional de Sitter spacetime, or rather, its double cover $\slalg$. For a self-contained review of the representation theory of $\slalg$, see \cref{app:rep}. 

{\it Stated concretely, our goal is to construct an $\slalg$-invariant interacting quantum mechanical system with states in the principal series.} To connect with de Sitter QFT, we would like such a model to have a tunable number $N$ of degrees of freedom (so that we may take a continuum limit), as well as a tunable interaction strength $\lambda$. 

A good place to start is the model introduced by Calogero in 1969 \cite{Calogero:1969xj}. The model is rather simple: it is composed of $N$ particles arranged on a line, and where each  pair of particles is coupled via a repulsive potential that decays with the square of the distance between them. Remarkably, this interacting model is exactly solvable for any $N$ and any value of the coupling strength $\lambda$. Despite its simplicity, particle scattering in this model displays evidence of anyonic statistics \cite{Polychronakos:1988tm}, which has garnered much interest in the condensed matter community due to its connections to the fractional quantum Hall effect \cite{Brink:1992xr,Brink:1993sz,Azuma:1993ra,Iso:1994ui,Polychronakos:1999sx,Ha:1994xj} and the matrix Chern-Simons theory \cite{Polychronakos:2001mi,Polychronakos:2001uw,Hellerman:2001rj}. We will give an extensive review of the Calogero model in the next section, particularly how to obtain its exact solution. But, for now, all we need to keep in mind is that it is an $\slalg$ invariant quantum mechanics with $N$ degrees of freedom and a tunable coupling $\lambda$. Unfortunately for us, the Hilbert space of the Calogero model is composed solely of  discrete series UIRs of $\slalg$, so is not suitable as a toy model of massive quantum fields in de Sitter. 

In this paper we will alter the Calogero model to skirt this obstruction. We take guidance from the following fact: the Calogero model happens to be a generalization of the De Alfaro-Fubini-Furlan (DFF) model \cite{deAlfaro:1976vlx}, a one-particle, $\slalg$-invariant quantum mechanics whose spectrum is a single discrete series module. Fortunately for us, in \cite{Andrzejewski:2011ya,Andrzejewski:2015jya}, the authors clarified how to modify the DFF model to accommodate principal series states, and this construction was later revisited in \cite{Anous:2020nxu}. We review this example in \cref{sec:n2princip} below and following it, we generalize this construction to the $N$-particle Calogero model. While we are clearly motivated by understanding solveable models of de Sitter QFT, in this paper we will content ourselves with setting the groundwork for establishing an interacting principal series system and leave any applications to de Sitter physics to future work. 

\subsection*{Primer on principal series vs other representations}
Before moving on, let us briefly remind the reader about the key differences between the various representations of $\slalg$, with more details provided in \cref{app:rep}, see also \cite{Anous:2020nxu,Anninos:2023lin,Sun:2021thf}. Recall that the algebra $\slalg$ can be expressed in a ladder basis $(l_0, l_\pm)$ satisfying:
\beq
[l_+,l_-]=2l_0~, \qquad\qquad [l_\pm,l_0]=\pm l_\pm ~.
\eeq
In this basis, the quadratic Casimir is expressed as: 
\beq
C_2=l_0^2-\frac{1}{2}\left(l_+l_-+l_-l_+\right)~,
\eeq
and we parametrize the Casimir eigenvalues as $\Delta(\Delta-1)$. First note that demanding that $\Delta(\Delta-1)\in \mathbb{R}$  implies either $\Delta\in \mathbb{R}$ or $\Delta\in\frac{1}{2}+i \mathbb{R}$, the latter case being the principal series. To gain slightly more intuition, note that we will take $l_0$ to be a compact generator, meaning its eigenvalues are integers, or half integers. Acting on an eigenbasis of $l_0$, it is straightforward to show, using the algebra above, that the generators act as follows: 
\beq\label{eq:lactionintro}
    l_0|r\rangle=-r|r\rangle~,\qquad l_\pm|r\rangle=-\left(r\pm \Delta\right)|r\pm 1\rangle~,\qquad r\in\mathbb Z~\text{or}~ 2r\in\mathbb Z~.
\eeq
Now observe the following, which is obvious from the above equation \eqref{eq:lactionintro}: if $\Delta\in\mathbb{Z}$ (or $2\Delta\in \mathbb Z$) then there exist a pair of states  $| r=\mp \Delta\rangle$, which are annihilated by $l_\pm$, respectively. These are a pair of highest (resp. lowest) weight states of the algebra, and we can generate the rest of the family by further action of $l_-$ (resp. $l_+$). This is the physics of the \emph{discrete series}, which behaves much like a harmonic oscillator and should be familiar from introductory texts on conformal field theory. There is a `ground state' which is annihilated by the appropriate operator, and a tower of excited states generated by the oscillator algebra. These can be thought of as bound states.\footnote{An incomplete list of papers discussing discrete series states in de Sitter include \cite{Anninos:2023lin,Letsios:2024snc,Letsios:2025pqo,RiosFukelman:2023mgq,Hinterbichler:2016fgl,Pethybridge:2024qci,Pethybridge:2021rwf,Joung:2007je,Loparco:2023akg,Loparco:2023rug}.} 

On the other hand, the remaining UIRs of $\slalg$, such as the principal series (with $\Delta\in \frac{1}{2}+i\mathbb R$), and complementary series ($\Delta\in (0,1)$), do not have a conventional `ground state.' They are never annihilated by any of the $l_\pm$. We will not have more to say on the complementary series here, but refer the reader to e.g. \cite{Sun:2021thf,Kitaev:2017hnr,Bargmann:1946me,gelfand1946unitary} if they are interested in learning more about this representation. The principal  series modules are unbounded, and akin to \emph{scattering states} or plane waves. These will be the focus of this paper. These states have made appearances in various different areas of theoretical physics lately, including in the context of black hole perturbations and near-extremal horizon instabilities \cite{Zimmerman:2016qtn, Gralla:2018xzo}, the AdS/CFT correspondence \cite{Kitaev:2018wpr, Gu:2019jub, Anninos:2019oka,Anninos:2017cnw}, in the decomposition of CFT correlators \cite{Caron-Huot:2017vep,Simmons-Duffin:2017nub,Kologlu:2019mfz,Kravchuk:2018htv}, in the celestial holography literature \cite{Pasterski:2017kqt,Donnay:2020guq,Atanasov:2021cje,Pasterski:2021raf}, and even more recently, to describe the chaotic dynamics near the spacelike singularity of a black hole \cite{Hartnoll:2025hly,DeClerck:2025mem}.

\section{Calogero model basics}\label{sec:Calobasics}

{We will begin by introducing the model and it solution as traditionally presented. While this section will serve primarily as a review, we will take care to present it in a manner we find physically intuitive while also establishing our notation.} The Calogero model \cite{Calogero:1969xj,Calogero:1970nt,Sutherland:1971ks,Moser:1975qp} is a quantum mechanics of $N$ particles arranged on a line, with a pairwise interaction that decays with the square of the inter-particle distance. Hence the model is defined by the following Hamiltonian: 
\beq \label{eq:calogeroH}
	H\equiv -\frac{1}{2}\sum_{i=1}^N\pa_i^2+\sum_{j<i}\frac{\lambda(\lambda-1)}{(x_i-x_j)^2}~.
\eeq
The parameter $\lambda$ is the coupling constant: in the limit $\lambda \to 0$ or $\lambda \to 1$ the system reduces to that of $N$ free particles.

An important feature of the Calogero model is that it defines a \emph{conformal} quantum mechanics \cite{Barucchi:1976am,Wojciechowski:1977uj}, which specifically means that there exist two additional operators, a generator of dilatations, $D$, and special conformal transformations, $K$, that, together with the Hamiltonian $H$, generate an $\slalg$ algebra. In the position basis the extra generators are as follows: 
\beq
	D\equiv -\frac{i}{4}\sum_{i=1}^N\left(x_i\pa_i+\pa_ix_i\right)~,\qquad K\equiv \frac{1}{2}\sum_{i=1}^Nx_i^2~, \label{eq:dkdefs}
\eeq
where it should be understood that the $\pa_i$'s appearing in $D$ act on all things to its right. It is easy to check that these three operators generate an $\slalg$ algebra:
\beq \label{eq:calogerosl2}
	[D,H]=iH~,\qquad [D,K]=-iK~,\qquad [K,H]=2iD~.
\eeq
As a consequence of this symmetry, the Hilbert space of the Calogero model is organized into unitary irreducible representations of $\slalg$. We provide a review of the representations of $\slalg$ in \cref{app:rep}.  However, let us state one of the main features here, as it will tie into the goal of this paper. Conformal families are classified according to their eigenvalues under the quadratic Casimir operator 
\beq
	C_2\equiv \frac{1}{2}\left(H\,K+K\,H\right)-D^2~,
\eeq
which on any irreducible representation takes the eigenvalue $C_2=\Delta(\Delta-1)$. Crucially, there are three families of unitary representations of $\slalg$: the \emph{discrete}, \emph{complementary} and \emph{principal series}. What distinguishes the principal series is that $\Delta$ takes complex values: $\Delta=\frac{1}{2}+i\nu$ for some $\nu\in\mathbb R$. Moreover, as written in \eqref{eq:calogeroH}, the Calogero Hamiltonian does not admit a Hermitian representation on the principal series for any real coupling $\lambda$, and to date, \emph{there is no known analytic continuation} of the Calogero model whose Hilbert space furnishes the principal series representation of $\slalg$. The goal of this paper is to fill this gap.

But before we get there, let us first review  the Calogero model as it is normally presented. Besides the $\slalg$ symmetry, the model is completely integrable. We will review this fact below and explain a simple procedure to generate the entire spectrum of the theory. With these facts in hand, we will specify how to analytically continue the model in order to obtain states in the principal series.

\subsection{Integrability of the Calogero model}
\subsubsection{Classical case}
The classical integrability of the Calogero model was first understood by Moser \cite{Moser:1975qp} and can be formulated as follows \cite{wojciechowski1983superintegrability,hikami1993classical}. Consider the classical Hamiltonian $\mathscr{H}$ (note the shift in the coupling constant as compared with the quantum case): 
\beq \label{eq:calogeroclassicalH}
	\mathscr{H}\equiv \frac{1}{2}\sum_{i=1}^Np_i^2+\sum_{j<i}\frac{\lambda^2}{(x_i-x_j)^2}~,
\eeq
where the set of $p_i$ and $x_j$ are canonically conjugate coordinates on phase space with $\{x_i,p_j\}=\delta_{ij}$
and where $\{\cdot,\cdot\}$ is the standard Poisson bracket. The classical equations of motion are then: 
\beq\label{eq:classicaleom}
\frac{\text{d} x_j }{\text{d} t}=\{x_j,\mathscr{H}\}=p_j~,\qquad\qquad \frac{\text{d} p_j }{\text{d} t}=\{p_j,\mathscr{H}\}=\sum_{k\neq j}^N\frac{2\lambda^2}{(x_j-x_k)^3}~.
\eeq
 It is possible to re-express the equations \eqref{eq:classicaleom} by defining a Lax pair of $N\times N$ matrices, $\mathscr{T}$ and $\mathscr{M}$, whose components are: 
\begin{align}
    \mathscr{T}_{jk}&\equiv \delta_{jk}\,p_j+i(1-\delta_{jk})\frac{\lambda}{x_j-x_k}~,\label{eq:laxclass} \\ \mathscr{M}_{jk}&\equiv -i(1-\delta_{jk})\frac{\lambda}{(x_j-x_k)^2}+\delta_{jk}\sum_{l\neq j}^N\frac{i\lambda}{(x_j-x_l)^2}~,
\end{align}
in terms of which \eqref{eq:classicaleom} can be recast as the following matrix equation:
\beq
\frac{\text{d}}{\text{d}t}\mathscr{T}=[\mathscr{T},\mathscr{M}]~.
\eeq
Moreover, this construction allows us to define a set of $N+1$ independent conserved quantities,\footnote{Admittedly, the current for $k=0$ is trivial, but we include it because it will be generated as the commutator of various non-trivial elements of the symmetry algebra.}\footnote{While nothing stops us from formally defining $\mathscr{J}_k$ for $k>N$, a simple consequence of the Cayley-Hamilton theorem is that these additional $\mathscr{J}_k$ are not independently conserved.} 
\beq 
\mathscr{J}_k\equiv \frac{1}{2}\,\text{Tr}\left(\mathscr{T}^k\right)~, \qquad k=0,1,\dots, N \label{eq:classicalconscurrents}
\eeq
whose equations of motion are \cite{wojciechowski1983superintegrability,hikami1993classical} 
\beq
\frac{\text{d} \mathscr{J}_k }{\text{d} t}=\left\{\mathscr{J}_k,\mathscr{H}\right\}=\frac{1}{2}\,\text{Tr}\,\left[\mathscr{T}^k, \mathscr{M}\right]=0~, 
\eeq
and hence they are all constants of motion with respect to the phase space flow generated by $\mathscr{H}$. Note moreover that we have: 
\beq
\mathscr{J}_2=\mathscr{H}~,
\eeq
so the Hamiltonian $\mathscr{H}$ itself sits in a classical current algebra generated by \eqref{eq:classicalconscurrents}.
The final step in proving the classical integrability of this system is to show that the following Poisson brackets vanish:
\beq
\left\lbrace\mathscr{J}_k,\mathscr{J}_l\right\rbrace=0~, \qquad \forall \, k,l
\eeq
as can be readily checked \cite{wojciechowski1983superintegrability,hikami1993classical}. This implies that one can choose any of the $\mathscr{J}_k$ to define a flow on phase space. 

\subsubsection{Quantum case}
To demonstrate that the quantum mechanical Calogero model (with Hamiltonian given in \eqref{eq:calogeroH}) remains integrable, one essentially follows the same steps as in the classical case (see e.g., \cite{Ujino1992,WADATI1993627,Ujino:1993gq}). Hence, one simply defines a quantum version of the Lax matrix $\mathscr{T}$  defined in \eqref{eq:laxclass} as follows: 
\beq
 \mathscr{T}_{jk}\rightarrow T_{jk}\equiv  -i\delta_{jk}\,\pa_j+i(1-\delta_{jk})\frac{\lambda}{x_j-x_k}~, \qquad j,k=1,\dots,N~.\label{eq:laxquant} 
\eeq
The matrix $T$ is now operator-valued and allows for a construction of the quantum analogs of the $N+1$ conserved currents \eqref{eq:classicalconscurrents}: 
\beq\label{eq:currentlax}
J_k\equiv \frac{1}{2}\,\sum_{i,j=1}^N\left(T^k\right)_{ij}~, \qquad k=0,\dots, N~.
\eeq
Note that, in contrast to the classical theory, the currents defined above are no longer constructed as \emph{traces} of powers of the $T$ matrix.  However, it still remains the case that only the first $N+1$ currents are linearly independent, as a consequence of the Cayley-Hamilton theorem. That is, the currents $J_{k>N}$ generated in this way can be expressed in terms of lower $J$'s (see e.g., \cite{Correa:2013rva}).  Nevertheless, as in the classical story, it turns out that these operators mutually commute
\begin{equation} \label{eq:commutingJ}
	[J_i,J_j]=0~,
\end{equation}
as was demonstrated in \cite{Ujino1992,WADATI1993627,Ujino:1993gq,Correa:2013rva}.
For concreteness, we list the first few currents explicitly
\begin{align}
	J_0&\equiv \frac{N}{2}~, \label{eq:J0}\\
	J_1&\equiv -\frac{i}{2}\sum_{i=1}^N \pa_i~,\\
	J_2&\equiv H~,\\
	J_3&\equiv \frac{i}{2}\left[\sum_{i=1}^N \pa_i^3-3\sum_{j<i}\frac{\lambda(\lambda-1)}{(x_i-x_j)^2}(\pa_i+
 \pa_j)\right]~, \label{eq:J3}\\
	&~~\vdots\nonumber
\end{align}
Note that, since $H$ is one of the currents, by virtue of \eqref{eq:commutingJ} all the other currents are conserved under the time evolution generated by $H$. 

\subsubsection*{\texorpdfstring{$\slalg$}{sl2alg} structure of the model}

We will now exploit the $\slalg$ structure of this model, which will allow us to construct the Hilbert space algebraically, in a way analogous (although complementary) to what was presented in \cite{Polychronakos:1992zk,Brink:1992xr,Brink:1993sz,Lapointe:1995ap}. Recall that our Hamiltonian $H$ lives in an $\slalg$ multiplet \eqref{eq:calogerosl2} 
\beq \label{eq:calogerosl22}
	[D,H]=iH~,\qquad [D,K]=-iK~,\qquad [K,H]=2iD~,
\eeq
with $D$ and $K$ defined in \eqref{eq:dkdefs}. The existence of these additional operators $D$ and $K$ are what singles out $J_2=H$ among the tower of currents.  We will exploit these extra operators to construct additional families of generators in this model. 

The first new family of generators is obtained by commuting the currents $J_k$ with the special conformal generator $K$, giving operators that we denote by $L_k$: 
\begin{equation} \label{eq:commutationKJ}
	[K, J_k]=\frac{k}{2}L_{k-2}~.
\end{equation}
Comparing \eqref{eq:commutationKJ} with \eqref{eq:calogerosl22}, we notice that $L_0=2iD$, meaning that, not only does the Hamiltonian $H$ sit in a multiplet with all the currents $J_k$, but $D$ sits in a multiplet with all the $L$'s.
It is a straightforward exercise to check that these new generators obey a Witt algebra under commutation
\begin{equation}\label{eq:Witt}
	[L_m,L_n]=(m-n)L_{m+n} ~. 
\end{equation}
For posterity, let us present explicit expressions for the first few of these generators: 
\begin{align}
	L_{-1}&\equiv i\sum_{i=1}^N x_i~,\\
	L_0&\equiv 2 i D~,\\
	L_1&\equiv 2J_1+i\left[-\sum_{i=1}^N x_i \pa_i^2+\sum_{j<i}\frac{\lambda(\lambda-1)}{(x_i-x_j)^2}(x_i+x_j)\right]~,\\
	L_2&\equiv 3 J_2+ \left[-\sum_{i=1}^N x_i \pa_i^3+\sum_{k<l}\frac{2\lambda(\lambda-1)}{(x_k-x_l)^2}\left(x_k\pa_k+x_l\pa_l+\frac{x_l\pa_k+x_k\pa_l}{2}+i\right)\right]~,\\
	&~~\vdots\nonumber
\end{align}
In fact, knowledge of these first few generators is enough to generate the remaining elements of the algebra by computing its closure. 

The index label $k$ denotes each operator's conformal dimension. To verify this, we check each operator's commutator with $L_0$---which, up to a constant multiple, is the same as its scaling dimension under the dilatation operator $D$---for example the $J_k$ and $L_k$ operators have:
\begin{equation}
[J_k,L_0]=k \,J_k~, \qquad\qquad [L_k,L_0]=k\, L_k ~. 
\end{equation}
From \eqref{eq:calogerosl22} it is then easy to see that $K$ has conformal dimension $-2$: 
\beq
[K,L_0]=-2K~,
\eeq
hence it lowers the $k$ label by two units, as in \eqref{eq:commutationKJ}.
Finally, the commutator algebra between these two multiplets is given by 
\begin{equation}\label{eq:U1current}
	[J_k,L_m]=k\, J_{k+m}~. 
\end{equation}
As we have already touched upon, only the $J_k$ and $L_k$ for $k\leq N$ span a set of linearly independent operators. However, as we can see from \eqref{eq:Witt} and \eqref{eq:U1current}, these linearly independent generators do not form a closed algebra, meaning we generate the universal enveloping algebra by successive commutation. Hence, as we take $N\rightarrow\infty$ we expect to find a conformal theory with a $U(1)$ current algebra structure, and for any finite $N$ we are working with a truncated set of currents.

\begin{figure}[t!]
\begin{center}
\includegraphics[width=0.35\paperwidth]{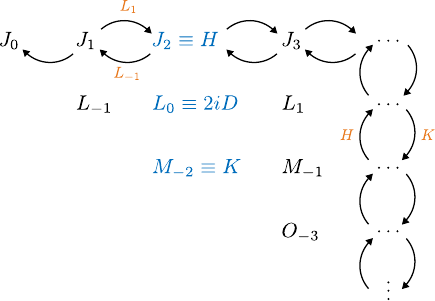}
\end{center}
\caption{The algebra of operators of the Calogero model. The top row denote the commuting currents in the current algebra, and the second row are the Witt algebra generators. The operators $H/K$ take us up/down a column, respectively, and the operators $L_1/L_{-1}$ move us right/left along a row, respectively. The column denoted in blue is special, as it is the only column whose operators form a closed subalgebra, in this case $\slalg$. Additionally, the operators in the first two rows form a closed subalgebra among themselves. }\label{fig:calogeroalgebra}
\end{figure}

It is useful to arrange the various operators in a table, as displayed in \cref{fig:calogeroalgebra}.
 The generators colored in blue span the $\slalg$ algebra while the generators in the first two rows fill out the current algebra. Commuting any operator with $K$ takes us down along the column in the table, while commuting with $H$ takes us up the column. As already demonstrated, the top row is annihilated by commutation with $H$. A column whose top component is $J_k$ has $k+1$ elements. Within a given row, action with $L_{\pm1}$ moves us to the right/left, respectively.

\subsection{Exact quantum spectrum: algebraic method}\label{sec:exact_quantum_spectrum_algebraic_method}

A common issue with $\slalg$-invariant quantum mechanics is that the Hamiltonian $H$ in \eqref{eq:calogeroH}, being a non-compact operator, has an unbounded spectrum, see for example \cite{deAlfaro:1976vlx,Kitaev:2017hnr}. This makes sense, the two-body interaction decays at large distances, so at best, the energy eigenstates will be plane-wave normalizable. The standard remedy for this is to work with the ladder-operator basis of the algebra:
\beq\label{eq:sl2algebrawithls}
l_0\equiv\frac{1}{2}\left(\frac{H}{\omega^2}+\omega^2 K\right)~,\qquad\qquad l_{\pm}\equiv \frac{1}{2}\left(\frac{H}{\omega^2}-\omega^2 K\right)\mp iD~,
\eeq
in terms of which the $\slalg$ algebra becomes:
\beq
[l_+,l_-]=2l_0~, \qquad\qquad [l_\pm,l_0]=\pm l_\pm ~.
\eeq
In this basis, the quadratic Casimir is expressed as: 
\beq
C_2=l_0^2-\frac{1}{2}\left(l_+l_-+l_-l_+\right)~.
\eeq
Going back to the definition \eqref{eq:dkdefs}, we see that working with $l_0$ amounts to adding a harmonic trap with characteristic frequency $\omega$ to the original Calogero Hamiltonian, hence the levels of the $l_0$ operator will be bounded and evenly spaced. We will henceforth set $\omega=1$, since we can reintroduce it at any time by a homogeneous rescaling of the $x_i$. From the algebra, it should be clear that $l_+$ lowers the $l_0$ eigenvalue by one unit, while $l_-$ raises it by one unit. 

Now comes the key insight that will allow us to algebraically solve for the spectrum of the Calogero model. Notice that it is possible to obtain the raising and lower operators $l_\pm$ from the Lax matrix operator \eqref{eq:laxquant}. To do this, first define a new pair of matrices $\tau_\pm$ whose components differ from $T$ only along the diagonal:
\beq
    (\tau_\pm)_{jk}\equiv T_{jk}\mp i\delta_{jk}x_j= -i\delta_{jk}\,(\pa_j\pm x_j)+i(1-\delta_{jk})\frac{\lambda}{x_j-x_k}~.
\eeq
In terms of these new matrices a quick calculation yields: 
\beq
l_{\pm}=\frac{1}{4}\sum_{i,j=1}^N\left(\tau^2_\pm\right)_{ij}~,
\eeq
which one should compare to \eqref{eq:currentlax}. Now recall that $l_-$ acts as a creation operator, whereas $l_+$ acts as an annihilation operator. 

These expressions lead us to consider the following set of $k=1,\dots,N$ generalizations of the above creation and annihilation operators: 
\begin{equation}\label{eq:creationannihilationdef}
a_k^\dagger\equiv -\left(\frac{i}{2}\right)^k\sum_{i,j=1}^N\left(\tau^k_-\right)_{ij}~,\qquad\qquad a_k\equiv -\left(-\frac{i}{2}\right)^k\sum_{i,j=1}^N\left(\tau^k_+\right)_{ij}~,
\end{equation}
from which we identify $a_2=l_+$ and $a^\dagger_2=l_-$.\footnote{Each $a_k$ and $a_k^\dagger$ can be written as a linear combination of operators in a single column from \cref{fig:calogeroalgebra}. For example: \begin{equation*} a_1= {i}\left(J_1-\frac{1}{2}L_{-1}\right)=\frac{1}{2}\sum_{i=1}^N\left(\pa_i+x_i\right)~,\qquad a^\dagger_1= -{i}\left(J_1+\frac{1}{2}L_{-1}\right)=-\frac{1}{2}\sum_{i=1}^N\left(\pa_i-x_i\right)~,\end{equation*} is a simple harmonic oscillator algebra for the center of mass of the $N$-Calogero particles. The choice to normalize these operators by powers of $(\pm i)^k$ in \eqref{eq:creationannihilationdef} {is a convention that} ensures that these operators {yield real wavefunctions} when acting on {(real)} bound-state wavefunctions. }
To convince the reader that these operators are indeed creation and annihilation operators with respect to the `Hamiltonian' $l_0$, it suffices to check that they satisfy: 
\beq \label{eq:koscillatoralgebra}
  \left[l_0,a_k\right]=-\frac{k}{2}a_k~,\,\qquad\qquad  \left[l_0,a^\dagger_k\right]=\frac{k}{2}a^\dagger_k~, \qquad\qquad [a_n,a_m]=\left[a_n^\dagger,a^\dagger_m\right]=0~,
\eeq
although note that 
\beq
\left[a_n^\dagger,a_m\right]\equiv B_{nm}\neq\delta_{mn}~.
\eeq
This is a key difference between the interacting model and a set of free harmonic oscillators. Regardless, as we will soon see, we will be able to build the Hilbert space of the interacting theory in a familiar way as the Fock space of $l_0$ eigenstates. The matrix $B_{mn}$ is complicated and we will not provide a general expression for its components, since it will not be needed to define this Fock space.

The Hilbert space of the Calogero model is constructed as follows. We first identify a ground state $|\Omega\rangle$ which satisfies: 
\beq
l_0|\Omega\rangle= \Delta_\Omega|\Omega\rangle~, \qquad a_k |\Omega\rangle=0 \quad \forall k~.
\eeq
The quantity $\Delta_\Omega$ denotes the state $|\Omega\rangle$'s eigenvalue under the quadratic Casimir,
\beq
    C_2|\Omega\rangle=\Delta_\Omega(\Delta_\Omega-1)|\Omega\rangle~,
\eeq
where the value of $\Delta_\Omega$ will be determined below. The Fock basis is then constructed as follows: Each state is labeled by a unique $N$-tuple of natural numbers $\{n_k\}=\{n_1,n_2,\dots,n_N\}$, which corresponds to the following wavefunction: 
\beq\label{eq:CalogeroFock}
    |\{n_k\}\rangle\equiv\frac{1}{\mathcal{N}_{\{n_k\}}}\prod_{k=1}^N\left(a^\dagger_k\right)^{n_k}|\Omega\rangle~,
\eeq
where $\mathcal{N}_{\{n_k\}}$ is an overall number which ensures that the state is normalized. Note that since the individual $a_j^\dagger$'s commute, their order does not matter in \eqref{eq:CalogeroFock}. A consequence of \eqref{eq:koscillatoralgebra} is that the the states $|\{n_k\}\rangle$ have $l_0$ eigenvalue: 
\beq
l_0|\{n_k\}\rangle=\left(\Delta_\Omega+\frac{1}{2}\sum_{k=1}^N k \,n_k\right)|\{n_k\}\rangle~.
\eeq
It is also interesting to write down the partition function of this model:
\beq\label{eq:partitionfunccalogero}
\boxed{\text{Tr}\, e^{-\beta\, l_0}=q^{2\Delta_\Omega}\prod_{k=1}^N\frac{1}{1-q^k}}~,\qquad\qquad q\equiv e^{-\frac{\beta}{2}}~,
\eeq
which we immediately recognize as a truncation of the partition function of a chiral-boson CFT \cite{DiFrancesco:1997nk}. As expected, the $N\rightarrow\infty$ limit of this model morphs into something resembling a local quantum field theory: the chiral half of a 1+1-$d$ Lorentz-invariant CFT with the structure of a $U(1)$ current algebra \cite{hikami1993classical,Cadoni:2000cz}.

\subsection{Solving for the ground state wavefunction}

\subsubsection{Center of mass and relative coordinates}

The final step is to determine the ground state wavefunction and its $l_0$ eigenvalue. To do this, we will find it useful to express the operators of the current algebra (most importantly the Hamiltonian \eqref{eq:calogeroH}) in terms of a new set of coordinates, which were introduced in the original paper \cite{Calogero:1970nt} but are standard (see e.g., \cite{OLSHANETSKY1983313}) and have appeared as recently as \cite{gonera1998calogero}. This new basis will allow us to determine $\Delta_\Omega$, as we will soon see. 

To this end, let us first separate out the center-of-mass variable
\beq \label{eq:comX}
X\equiv \frac{1}{{N}}\sum_{i=1}^Nx_i~.
\eeq
We also introduce a special set of relative coordinates (whose usefuleness will become clear as we progress)
\beq\label{eq:relativcoorddef}
	y_k\equiv \sqrt{\frac{k}{k+1}}\left(\,x_{k+1}-\frac{1}{k}\sum_{j=1}^k x_j\right)~,\qquad 1\leq k\leq N-1~,
\eeq 
and let us set $y_N\equiv {\sqrt{N}} X$. Importantly, the change of basis from $\{x_k\}$ to $\{y_k\}$ is orthogonal. Indeed, if we introduce a matrix $A$ that implements the inverse change of basis 
\beq \label{eq:basischangey}
x_i=\sum_{k=1}^NA_{ik}y_{k}~,\qquad 
\eeq
it is easy to check that $A^{T}A=AA^T=\mathds{1}$ (see \cref{app:coordtrafo}). Focusing on the $J_1$ and $J_2$ columns in \cref{fig:calogeroalgebra}, we notice that the operators in these columns obey a nice splitting in terms of the center-of-mass and relative coordinate. For instance, it is straightforward to see that, in these coordinates, the $J_1$ column only acts on the center of mass:  
\begin{equation}
J_1=-\frac{i}{2}\partial_X~,\qquad L_{-1}=i N\,X~,\qquad a_1=\frac{1}{2}\left(\partial_X+N X\right),\qquad a^\dagger_1=-\frac{1}{2}\left(\partial_X-N X\right)~.
\end{equation}
Moreover, from the orthogonality of the basis transformation, it immediately follows that the generators $K$ and $D$ can be written as follows:  
\begin{equation}\label{eq:KDreldecomp}
	K = K_X+K_{\rm rel}~,\qquad D = D_X+D_{\rm rel}~,
\end{equation}
where 
\begin{align}
K_X &\equiv \frac{N}{2}X^2~, &D_X=-\frac{i}{4}\left(\partial_X\,X+X\,\partial_X\right)~,\\ K_{\rm rel}&\equiv \frac{1}{2}\mathbf{y}^2~,  &D_{\rm rel}\equiv -\frac{i}{4}\left(\partial_{\mathbf{y}}\cdot\mathbf{y}+\mathbf{y}\cdot\partial_{\mathbf{y}}\right)~,
\end{align}
where we have defined an $(N-1)$-vector $\mathbf{y}\equiv(y_1,\ldots, y_{N-1})$ and, in the above expression, the dot denotes the standard Euclidean product on $\mathbb{R}^{N-1}$ (and hence $\mathbf{y}^2=\mathbf{y}\cdot\mathbf{y}$).

The Hamiltonian also admits a simple expression in these variables, owing to the fact that the the potential only depends on position \emph{differences}, meaning the center-of-mass dependence drops out:
\beq 
x_i-x_j = \sum_{k=1}^{N}(A_{ik}-A_{jk}) y_{k}=\sum_{k=1}^{N-1}(A_{ik}-A_{jk}) y_{k}=\sqrt{2}\sum_{k=1}^{N-1} b^{[i,j]}_k y_k ~, 
\eeq
where we have defined the coefficients $b^{[i,j]}_k$ as follows:
\beq \label{eq:ba}
b^{[i,j]}_k\equiv \frac{1}{\sqrt{2}}\left(A_{ik}-A_{jk}\right)~.
\eeq
From here on, it will be useful to introduce a multi-index, denoted by $a=[i,j]$, with $i>j$, which runs over all the $N(N-1)/2$ independent choices of a particular pair of particles, and collect the above coefficients into a set of $N(N-1)/2$ vectors $\mathbf{b}^a=(b^a_1,\ldots, b^a_{N-1})\in \mathbb{R}^{N-1}$. Now we can see that, like \eqref{eq:KDreldecomp}, the Hamiltonian also admits a center-of-mass and relative decomposition of the form:
\beq\label{eq:Hamiltoniandecomposed}
	H=H_X+H_{\rm rel}~,
\eeq
where 
\beq \label{eq:relativeHamiltonian}
H_X\equiv-\frac{1}{2N}\pa_X^2~,\qquad H_{\rm rel}\equiv -\frac{1}{2}\pa_{\mathbf{y}}^2 +\frac{1}{2}\sum_{a}\frac{\lambda(\lambda-1)}{\left(\mathbf{b}^a\cdot \mathbf{y}\right)^2}~,
\eeq
and, in the above expression, the dot again denotes the standard Euclidean product on $\mathbb{R}^{N-1}$:
\beq \label{eq:innerproductby}
\mathbf{b}^a\cdot \mathbf{y} \equiv \sum_{k=1}^{N-1} b_{k}^a y_k~,
\eeq
and $\pa_{\mathbf{y}}^2\equiv \pa_{\mathbf{y}}\cdot \pa_{\mathbf{y}}$~.
An important feature of the $\mathbf{b}^a$'s is that they are unit vectors with respect to the above inner product. This follows from the orthogonality of the matrix $A$. To be precise, we have
\beq
	\mathbf{b}^a\cdot \mathbf{b}^{c}= \sum_{k=1}^{N-1}b_k^ab_k^c=\gamma^{ac}~,
\eeq
where the multi-index matrix $\gamma^{ab}$ is straightforward to compute:
\beq\label{eq:borth}
	\gamma^{[ij],[i'j']}\equiv \frac{1}{2}\left(\delta_{ii'}-\delta_{ij'}+\delta_{jj'}-\delta_{ji'}\right)~.
\eeq
In fact, this shows that the $\mathbf{b}^a$ correspond precisely to the (normalized) positive roots of the Lie algebra $A_{N-1}=\mathfrak{su}(N)$. In particular, this provides us with an explicit expression for the angles between the $N(N-1)/2$ unit vectors $\mathbf{b}^a$. 

\subsubsection{Radial and angular variables}

Let us now introduce spherical coordinates on the vector $\mathbf{y}$ in $\mathbb{R}^{N-1}$. We first introduce the radial coordinate in the usual way:
\beq \label{eq:rdef}
r^2 \equiv\mathbf{y}\cdot\mathbf{y}=\frac{1}{N}\sum_{j<i}(x_i-x_j)^2~,
\eeq
and angular variables $\omega_i$, $i=1,\ldots, N-1$, which parameterize a unit $(N-2)$-sphere via
\begin{align}
	\omega_1&=\cos\theta_1~,\nonumber\\
	\omega_2&=\sin\theta_1\,\cos\theta_2~,\nonumber\\
	\vdots&\nonumber\\
	\omega_{N-2}&=\sin\theta_1\ldots\sin\theta_{N-3}\cos\theta_{N-2}~,\nonumber\\
	\omega_{N-1}&=\sin\theta_1\ldots\sin\theta_{N-3}\sin\theta_{N-2}~.\label{eq:nm2sphere}
\end{align}
Under the change of variables $y_i =r\omega_i$, a standard calculation reveals that the relative operators become:
\begin{align}
 H_{\rm rel}&=-\frac{1}{2\,r^{N-2}}\partial_r\left( r^{N-2}\,\partial_r\right)+\frac{1}{2r^2}\left(-\nabla^2_{S^{N-2}}+\sum_{a}\frac{\lambda(\lambda-1)}{(\cos\Theta^a)^2}\right)~,\label{eq:gensphericalcoord} \\ K_{\rm rel}&=\frac{r^2}{2}~,\label{eq:gensphericalcoordKrel} \\ D_{\rm rel}&=-\frac{i}{4}\left(r^{2-N}\partial_r\, r^{N-1}+r\,\partial_r\right)~,\label{eq:gensphericalcoordDrel}
\end{align}
where $\nabla^2_{S^{N-2}}$ is the Laplacian on the unit $S^{N-2}$, and we have introduced the notation $\Theta^a$ for the angle between the unit vector $\mathbf{b}^a$ and the vector $\mathbf{y}$, as computed by the inner product in \eqref{eq:innerproductby}:
\beq\label{eq:Thetaadef}
	\mathbf{b}^a\cdot \mathbf{y}=\sum_{i=1}^{N-1}b^a_iy_i\equiv r\,\cos\Theta^a~.
\eeq
In other words, we see that there is a clean split between the radial and angular dynamics:
\beq\label{eq:relativeHampolar}
{H}_{\rm rel}=-\frac{1}{2}r^{2-N}\pa_r\left(r^{N-2}\,\pa_r\right)+\frac{1}{2r^2}\hat{L}^2_{S^{N-2}}~,
\eeq
where we have defined the following `angular' operator: 
\beq \label{eq:angularpart}
 \hat{L}^2_{S^{N-2}}\equiv-\nabla^2_{S^{N-2}}+\sum_{a}\frac{\lambda(\lambda-1)}{(\cos\Theta^a)^2}~.
\eeq
Finding the ground state of the Calogero model also means finding the ground state of $\hat{L}^2_{S^{N-2}}$. Luckily, there is a convenient trick that we can use, as described in the next section.

\subsubsection{A similarity transformation}
The above expression for the Hamiltonian \eqref{eq:Hamiltoniandecomposed} can be simplified by a judicious choice of similarity transformation, as we will now show. Let us first introduce the following scalar:
\beq\label{eq:vanddef}
	\vdm(\mathbf{y})\equiv \frac{\prod_{j<i}(x_i-x_j)}{\left(2r^2\right)^{\frac{N(N-1)}{4}}}~,
\eeq
which, up to an overall constant, is the usual determinant of the Vandermonde matrix built out of the $x_i$'s, rescaled by a power of the relative radial coordinate $r$. Using \eqref{eq:basischangey} we find:
\beq 
x_i-x_j = \sum_{k=1}^{N-1}(A_{ik}-A_{jk}) y_{k} = \sqrt{2}\sum_{k=1}^{N-1} b^{[i,j]}_k y_k =\sqrt{2}\,r\,\cos\Theta^a~,
\eeq
hence $\vdm(\mathbf{y})$ only depends on the angles $\Theta^a$ defined in \eqref{eq:Thetaadef}:
\beq \label{eq:delta}
\vdm(\mathbf{y}) =\prod_{j<i}\left(\frac{\mathbf{b}^{[i,j]}\cdot \mathbf{y}}{r}\right)= \prod_a\,\cos\Theta^a~.
\eeq
Now comes the trick: We will denote by a caligraphic script, any operator related to our original algebra by the following similarity transformation, e.g.,
\begin{equation}\label{eq:similarity}
\mathcal{H}\equiv \vdm^{-\lambda} H\vdm^\lambda~, \qquad \mathcal{D}\equiv \vdm^{-\lambda} D\vdm^\lambda~,\qquad \mathcal{K}\equiv \vdm^{-\lambda} K\vdm^\lambda~.
\end{equation}
Note that since $K$ and $D$ only act on functions of $X$ and $r$, but are otherwise independent of the angles $\Theta^a$ (see equations \eqref{eq:KDreldecomp} and (\ref{eq:gensphericalcoordKrel}-\ref{eq:gensphericalcoordDrel})), we have that 
\beq
\mathcal{D}=D~,\qquad \mathcal{K}=K~.
\eeq
Note also that, starting from any  eigenfunction $\Psi=\Psi(X,\mathbf{y})$ of an arbitrarily-chosen caligraphic operator $\mathcal{O}$ satisfying $\mathcal{O}\Psi=E \Psi$, we can construct an eigenfunction of $O$ by noting that $O\vdm^\lambda \Psi = E\vdm^\lambda \Psi$. Hence the spectra of the caligraphic operators are the same as those of the original operators, as is always the case with a similarity transformation.  

Now note that for a test function $f=f(X,\mathbf{y})$ we have
\begin{align} \label{eq:conjugationH}
	\mathcal{H}(f)\equiv\vdm^{-\lambda} H\left(\vdm^\lambda f\right)=&-\frac{1}{2N}\pa_X^2f-\frac{1}{2}\pa_{\mathbf{y}}^2f-\frac{\lambda}{\vdm}\left(\pa_{\mathbf{y}}\vdm\right)\cdot\left(\pa_{\mathbf{y}}f\right)\nonumber\\
	&-\frac{\lambda}{2}\left(\frac{\pa^2_{\mathbf{y}}\vdm}{\vdm}+(\lambda-1)\left[\frac{(\pa_{\mathbf{y}}\vdm)\cdot (\pa_{\mathbf{y}}\vdm)}{\vdm^2}-\sum_a\frac{1}{(\mathbf{b}^a\cdot \mathbf{y})^2}\right]\right)\,f~.
\end{align}
We will now show that the coefficient of $f$ in the second line in \eqref{eq:conjugationH} is a spherically symmetric function, meaning it depends only on the radial variable $r$. 
First note, using \eqref{eq:delta} and \eqref{eq:rdef}, that
\begin{equation}
\log \vdm(\mathbf{y}) =-\frac{N(N-1)}{4}\log\left({\mathbf{y}\cdot\mathbf{y}}\right)+\sum_{c}\log\left(\mathbf{b}^c\cdot \mathbf{y}\right)~.
\end{equation}
Taking derivatives of the above expression, we find (no sum on $i$):
\begin{align}
\pa_{\mathbf{y}_i}\log\vdm=&\frac{\pa_{y_i}\vdm}{\vdm}&=&\left(-\frac{N(N-1)}{2}\frac{y_i}{r^2}+\sum_c\frac{b^c_i}{\mathbf{b}^c\cdot \mathbf{y}}\right)~,\label{eq:firstderivlog}\\
\pa_{\mathbf{y}_i}^2\log\vdm=&\frac{\pa_{y_i}^2\vdm}{\vdm}-\frac{(\pa_{y_i}\vdm)^2}{\vdm^2}&=&\left(-\frac{N(N-1)}{2r^2}+N(N-1)\frac{(y_i)^2}{r^4}-\sum_c\frac{(b^c_i)^2}{(\mathbf{b}^c\cdot \mathbf{y})^2}\right)~.\label{eq:secondderivlog}
\end{align}
By dotting \eqref{eq:firstderivlog} into itself, and using the known expression for the angles between the $\mathbf{b}^c$ vectors \eqref{eq:borth}, we have: 
\begin{equation}
\frac{(\pa_{\mathbf{y}}\vdm)\cdot (\pa_{\mathbf{y}}\vdm)}{\vdm^2}=\left(-\frac{N^2(N-1)^2}{4r^2}+\sum_c\frac{1}{(\mathbf{b}^c\cdot \mathbf{y})^2}\right) ~. \label{eq:sumDelta1}
\end{equation}
Summing equation \eqref{eq:secondderivlog} over $i$ and adding \eqref{eq:sumDelta1} to it, we also find: 
\begin{equation}
\frac{\pa^2_{\mathbf{y}}\vdm}{\vdm}=-\frac{N(N-1)}{2r^2}\left(\frac{N(N-1)}{2}+N-3\right)~. \label{eq:sumDelta2}
\end{equation}
Putting everything together, one obtains the following result for the conjugated Hamiltonian:
\beq
	\boxed{\mc H=-\frac{1}{2N}\pa_X^2-\frac{1}{2\vdm^{2\lambda}}\pa_{\mathbf{y}}\cdot\left(\vdm^{2\lambda}\pa_\mathbf{y}\right)+\lambda\frac{N(N-1)}{4r^2}\left(\lambda\frac{N(N-1)}{2}+N-3\right)}~, 
\eeq
where one can check that this operator (and the all the other conjugated operators) are Hermitian with respect to the inner product:\footnote{Care must be taken when defining the limits of $\mathbf{y}$ integrals, as we will see in the explicit solution of the $N=3$ problem. }
\beq
(f,g)=\int dX\int  d^{N-1}\mathbf{y}\,\vdm^{2\lambda}f^*\,g~.
\eeq
Now, keeping in mind that $\vdm$ only depends on the angles $\Theta^a$, and that we have a splitting
\beq
\mathcal{H}=\mathcal{H}_X+\mathcal{H}_{\rm rel}~,
\eeq
and that the original relative Hamiltonian had a radial and angular decomposition as in \eqref{eq:relativeHampolar} and \eqref{eq:angularpart}, we find that:
\beq
	\mc{H}_{\rm rel}=-\frac{1}{2}r^{2-N}\pa_r\left(r^{N-2}\pa_r\right)+\frac{1}{2r^2}\hat{\mc{L}}^2_{S^{N-2}}~,
\eeq
where the new angular operator $\hat{\mc{L}}^2_{S^{N-2}}$ is defined as follows:
\beq \label{eq:reducedangH}
	\hat{\mc{L}}^2_{S^{N-2}}
\equiv -\frac{\vdm^{-2\lambda}}{\sqrt{g}}\pa_\mu\left(\vdm^{2\lambda}\sqrt{g}g^{\mu\nu}\pa_\nu\right)+\lambda\frac{N(N-1)}{2}\left(\lambda\frac{N(N-1)}{2}+N-3\right)~,
\eeq
where $g_{\mu\nu}$ is the round metric on the unit sphere $S^{N-2}$. The above manipulations emphasize that the angular differential operator, in this conjugated frame, is simply a linear deformation of the spherical Laplacian: 
\beq \label{eq:Mhat}
	-\frac{\vdm^{-2\lambda}}{\sqrt{g}}\pa_\mu\left(\vdm^{2\lambda}\sqrt{g}g^{\mu\nu}\pa_\nu\right)=-\nabla^2_{S^{N-2}}-\lambda\hat M~,
\eeq
where $\hat M$ is a first-order differential operator on $S^{N-2}$. Hence, the caligraphic angular part of the Hamiltonian is of the form  
\beq \label{eq:reducedangH2}
	\hat{\mc{L}}^2_{S^{N-2}}=-\nabla^2_{S^{N-2}}-\lambda\hat M+\lambda\frac{N(N-1)}{2}\left(\lambda\frac{N(N-1)}{2}+N-3\right)~,
\eeq
It is clear that in the limit $\lambda \to 0$ the above expression reduces to the usual spherical Laplacian.

\paragraph{Summary:}
We have shown that our conjugated $\slalg$ operators admit the following split in terms of center-of-mass and relative coordinates: 
\beq\label{eq:fullconjugated}
\mc{H}= \mc{H}_X+\mc{H}_{\rm rel}~, \qquad \mc K = \mc{K}_X+\mc{K}_{\rm rel}~,\qquad \mc D= \mc{D}_X+\mc{D}_{\rm rel}~,
\eeq
where 
\beq\label{eq:conjugatedX}
\mathcal{H}_X=-\frac{1}{2N}\pa_X^2~, \qquad \mc{K}_X=  \frac{N}{2}X^2~,\qquad \mc{D}_X=-\frac{i}{4}\left(\partial_X\,X+X\,\partial_X\right)~.
\eeq
The center-of-mass degree of freedom forms a closed $\slalg$ algebra with Casimir: 
\beq
	\mc{C}_2^X\equiv \frac{1}{2}\left(\mc{H}_X\,\mc{K}_X+\mc{K}_X\,\mc{H}_X\right)-\mc{D}_X^2=-\frac{3}{16}=\frac{1}{4}\left(\frac{1}{4}-1\right)~,
\eeq
where we immediately recognize that the conformal weight of the center-of-mass multiplet is $\Delta_X\equiv\frac{1}{4}$.
Similarly, we found
\begin{multline}\label{eq:conjugatedrel}
\mc{H}_{\rm rel}=-\frac{1}{2\,r^{N-2}}\partial_r\left( r^{N-2}\,\partial_r\right)+\frac{1}{2r^2}\hat{\mc{L}}^2_{S^{N-2}}~,\\ \mc{K}_{\rm rel}=\frac{r^2}{2}~,\qquad \mc{D}_{\rm rel}=-\frac{i}{4}\left(r^{2-N}\partial_r\, r^{N-1}+r\,\partial_r\right)~,
\end{multline}
which also forms a closed $\slalg$ algebra with Casimir: 
\beq\label{eq:relcasimir}
	\mc{C}_2^{\rm rel}\equiv \frac{1}{2}\left(\mc{H}_{\rm rel}\,\mc{K}_{\rm rel}+\mc{K}_{\rm rel}\,\mc{H}_{\rm rel}\right)-\mc{D}_{\rm rel}^2=\frac{1}{4}\left(\hat{\mc{L}}^2_{S^{N-2}}+\frac{(N-5)(N-1)}{4}\right)~,
\eeq
hence the value of the casimir $\Delta_{\rm rel}$ will be determined by the wavefunction's eigenvalue under the angular operator $\hat{\mc{L}}^2_{S^{N-2}}$~. Also note that since we are ultimately interested in the spectra of the full operators \eqref{eq:fullconjugated}, apart from the ground state,\footnote{This follows from writing the Casimir as $\mc{C}_2=\ell_0(\ell_0-1)+\ell_-\ell_+$ with $\ell_+$ annihilating the ground state.} the eigenvalues of the full Casimir will not be as simple as adding $\Delta_X+\Delta_{\rm rel}=\frac{1}{4}+\Delta_{\rm rel}$. This much should be clear to any undergraduate who has struggled with Clebsch-Gordan coefficients while learning how to add angular momenta.

\subsubsection{The ground state wavefunction}
We are finally ready to solve for the ground state wavefunction of the Calogero model as well its eigenvalue $\Delta_\Omega$. From it, as we described in \cref{sec:exact_quantum_spectrum_algebraic_method}, we can generate the entire spectrum by acting via the action of the creation operators. From the definition of the angular operator $\hat{\mc{L}}^2_{S^{N-2}}$ \eqref{eq:reducedangH}, it is not unreasonable to expect the ground state to be a constant function over the $S^{N-2}$. Moreover, since the operators in the $J_1$ and $J_2$ columns split nicely between center-of-mass degrees of freedom $X$, and relative coordinats $\mathbf{y}$, we expect the ground state wavefunction (of the conjugated operators) to be of the form:\footnote{It is also reasonable to predict that the excited states will not split nicely in this way, by action of the higher creation operators, which mix between $X$ and $\mathbf{y}$.} 
\beq
\Psi_\Omega=\chi(X)\rho(r)~.
\eeq
Already at this stage it is possible for us to  determine $\Delta_\Omega$ without the need to solve a differential equation. Let us explain how. On wavefunctions that are constant over the $S^{N-2}$, the operator $\hat{\mc{L}}^2_{S^{N-2}}$ is a constant: 
\beq
\hat{\mc{L}}^2_{S^{N-2}}\Psi_\Omega=\lambda\frac{N(N-1)}{2}\left(\lambda\frac{N(N-1)}{2}+N-3\right)\Psi_\Omega~,
\eeq
hence we can immediately deduce from \eqref{eq:relcasimir} that: 
\beq
\mc{C}_2^{\rm rel}\Psi_\Omega=\Delta^\Omega_{\rm rel}(\Delta^\Omega_{\rm rel}-1)\Psi_\Omega~,
\eeq
with\footnote{A nice consistency check is that $\lim_{\lambda\rightarrow0}\Delta_{\rm rel}^\Omega=\frac{N-1}{4}=(N-1)\Delta_X$, the conformal dimension of $N-1$ free particles. }
\beq\label{eq:deltaomegarelanyN}
\Delta^\Omega_{\rm rel}\equiv \frac{(N\lambda+1)(N-1)}{4}~.
\eeq
For the ground state, we can add  {together} the center {of} mass {and the relative} Casimirs and we deduce $\Delta_\Omega$:
\beq\label{eq:DeltaOmegaboxed}
\boxed{\Delta_\Omega=\Delta_X+\Delta_{\rm rel}^\Omega=\frac{\lambda N (N-1)+N}{4}}~.
\eeq 
This completes our calculation of the partition function \eqref{eq:partitionfunccalogero}. We reiterate that this model has the same partition function as a truncated chiral boson, where the ground state energy $\Delta_\Omega$ naively diverges in the continuum $N\rightarrow\infty$ limit. This is a standard divergence that one encounters when taking the field theory limit. 

Given our ansatz for $\Psi_\Omega$, finding the ground state involves solving the following two differential equations: 
\begin{equation}
a_1 \chi(X)=\frac{1}{2}\left(\pa_X+NX\right)\chi(X)=0~,
\end{equation}
and
\begin{multline}
\ell_+^{\rm rel} \rho(r)=-\frac{1}{4}\left(\frac{1}{r^{N-2}}\pa_r\left(r^{N-2}\pa_r\,\rho(r)\right)+2r\pa_r\rho(r)\right.\\\left.-\left[\lambda\frac{N(N-1)}{2r^2}\left(\lambda\frac{N(N-1)}{2}+N-3\right)-r^2-(N-1)\right]\rho(r)\right)=0~.
\end{multline}
One can easily check that this leads to: 
\beq
\Psi_\Omega(X,r)=\frac{\left(r^2\right)^{\Delta_\Omega-\frac{N}{4}}}{\mathcal{N}}e^{-\frac{N}{2}X^2-\frac{r^2}{2}}~, 
\eeq
where $\mathcal{N}$ is a normalization constant. This allows us to go back and write down the ground state of the unconjugated operators in the original variables $x_i$. Defining: 
\beq
\psi_\Omega(\vec{x})=\frac{\prod_{j<i}(x_i-x_j)^\lambda}{\mathcal{N}}e^{-\frac{1}{2}\sum_i x_i^2}~,
\eeq
then one can check that it satisfies: 
\beq
l_0\psi_\Omega=\Delta_\Omega \psi_\Omega~.
\eeq
Subsequently the Hilbert space will consists of states
\beq
\psi_{\{n_k\}}\equiv\frac{1}{\mathcal{N}_{\{n_k\}}}\prod_{k=1}^N\left(a^\dagger_k\right)^{n_k}\psi_\Omega~,
\eeq
which statisfy
\beq\label{eq:spectrumagain}
l_0\psi_{\{n_k\}}=\left(\Delta_\Omega+\frac{1}{2}\sum_{k=1}^N k \,n_k\right)\psi_{\{n_k\}}~.
\eeq

One final comment is in order. We have studied an \emph{interacting}, multi-particle quantum mechanics with coupling constant $\lambda(\lambda-1)$. Now, granted, this model has a lot of symmetry, but the truly remarkable feature is that the spectrum is exactly \emph{linear} in  $\lambda$ for all values of the coupling. This is very different from the expectation from perturbation theory: linear deformations of the Hamiltonian typically lead to deformations of the spectrum at all orders. This truncation of perturbation theory is the result of integrability.

\section{The principal series Calogero model} 

Having outlined the procedure for solving the Calogero model for any $N$ and $\lambda$, we will now describe what is needed to analytically continue these models so that their spectra accommodate the principal series of $\slalg$. The first step in this endeavor is to focus only on the relative system. {We will see that the relative system is sufficient to realize our goal of constructing interacting principal series states.} That means that, going forward, we will ignore all dependence on the center-of-mass variable $X$, and set it to zero everywhere. {However as illustrated in the solution for the standard Calogero model, the center of mass remains an important degree of freedom for the physical interpretation of the model, especially in the large-$N$ limit. We will not attempt to give suitable prescription for continuing the center of mass at this time, although we expect it to play a similarly important role in our continued model (the reader can see \cref{sec:disc} for further comments). The reason being that, while the center-of-mass multiplet is in the discrete series representation of $\slalg$, the fact that $\Delta_X=\tfrac{1}{4}$ in the Calogero model implies that the symmetry algebra of standard Calogero model actually corresponds to the universal cover  of the algebra $\widehat{\slalg}$. For the principal series we will keep the global structure of $\slalg$ rather than its universal cover. It would be interesting to explore what adding a single discrete, principal, or complementary series center of mass multiplet would achieve in this case, which we leave to future work. }

\subsection{\texorpdfstring{$N=2$}{N=2}}
We will first present the $N=2$ case, which is none other than the De Alfaro-Fubini-Furlan (DFF) model \cite{deAlfaro:1976vlx}, whose analytic continuation to the principal series was explained in \cite{Andrzejewski:2011ya,Andrzejewski:2015jya} and later reviewed in \cite{Anous:2020nxu}. 
\subsubsection{Setup and original solution}
For $N=2$, before analytic continuation, we have the following operators (c.f. \eqref{eq:conjugatedrel}):
\begin{equation}\label{eq:conjugatedreln2}
\mc{H}_{\rm rel}=-\frac{1}{2}\partial_r^2+\frac{\lambda(\lambda-1)}{2r^2}~,\qquad\mc{K}_{\rm rel}=\frac{r^2}{2}~,\qquad \mc{D}_{\rm rel}=-\frac{i}{4}\left(\partial_r\, r+r\,\partial_r\right)~,
\end{equation}
along with the quadratic Casimir, which evaluates to a constant: 
\beq
\mc{C}_2^{\rm rel}=\frac{(2\lambda+1)(2\lambda-3)}{16}~,
\eeq
and indicates that the spectrum corresponds to a single $\Delta=\frac{1}{4}+\frac{\lambda}{2}$ conformal multiplet (see \eqref{eq:deltaomegarelanyN} for $N=2$). The spectrum of $\ell_0^{\rm rel}$ eigenstates, where
\begin{equation}
	\ell_0^{\rm rel}= \frac{1}{2}\left(\mc{H}_{\rm rel}+\mc{K}_{\rm rel}\right)~,
\end{equation}
are straightforward to express in terms of generalized Laguerre polynomials $L_k^\alpha$ for $k=0,1,2,\dots$:
\begin{equation} \label{eq:Laguerresolutions}
	\eta_k(r)\equiv \sqrt{\frac{2(k!)}{\Gamma(k+2\Delta)}}e^{-\frac{r^2}{2}}r^{2\Delta-\frac{1}{2}}L_k^{2\Delta-1}\left(r^2\right)~.
\end{equation}
In particular, they satisfy the eigenvalue equation
\begin{equation}
	\ell_0^{\rm rel}\eta_k(r)=(k+\Delta)\eta_k(r)~,
\end{equation}
and are orthonormal with respect to the norm $(\cdot,\cdot)$ 
\begin{equation}\label{eq:normn2}
	(f,\,g)\equiv\int_0^\infty \dd r\, f^*(r)\,g(r)~. 
\end{equation}

\subsubsection{Analytic continuation to the principal series}\label{sec:n2princip}
In order to analytically continue this model to the principal series, we first need to take $\Delta=\frac{1}{2}+i\nu$, which translates to continuing the parameter $\lambda\rightarrow\frac{1}{2}+2i\nu$. Plugging this into \eqref{eq:conjugatedreln2}, we find that this analytic continuation corresponds to an \emph{attractive} $1/r^2$ potential, rather than a repulsive one. But a quantum mechanics with an attractive $1/r^2$ potential suffers from the famous `falling into the origin' problem \cite{landau2013quantum}, owing to the fact that the Hamiltonian fails to be self-adjoint with respect to the norm \eqref{eq:normn2}. So it is not sufficient to simply continue the parameter $\lambda$ in order to furnish states in the principal series. 

As explained in \cite{Andrzejewski:2011ya,Andrzejewski:2015jya,Anous:2020nxu}, the issue at hand is that the representation of the operators \eqref{eq:conjugatedreln2} forces $\mathcal{K}_{\rm rel}$ to have non-negative eigenvalues, which is the main tension with the principal series. To remedy this, in addition to continuing the parameter $\lambda$, we continue $r\rightarrow \sqrt{2\kappa}$ and extend the range of $\kappa$ to $\kappa\in(-\infty,\infty)$. { The result of this continuation is:}
\begin{equation}\label{eq:conjugatedkappan2}
\mc{H}_{\rm rel}=-\kappa\partial_{\kappa}^2-\frac{1}{2}\partial_\kappa+\frac{\left(\Delta-\frac{3}{4}\right)\left(\Delta-\frac{1}{4}\right)}{\kappa}~,\qquad\mc{K}_{\rm rel}=\kappa~,\qquad \mc{D}_{\rm rel}=-i\left(\kappa\partial_\kappa+\frac{1}{4}\right)~,
\end{equation}   
with $\Delta=\frac{1}{2}+i\nu$, which one can check properly furnishes the algebra \eqref{eq:calogerosl2}. {Defining $\kappa\equiv\text{sgn}(r)\tfrac{r^2}{2}$, notice that the relative Hamiltonian switches sign across $r=0$ which is reminiscent of the flip of the sign of the Rindler Hamiltonian in the Kruskal extension across a horizon, although it is hard to go beyond analogizing at this stage.} The operators \eqref{eq:conjugatedkappan2} are Hermitian with respect to the inner product $(\cdot,\cdot)$ 
\begin{equation}\label{eq:normn2kappa}
	(f,\,g)\equiv\int_{-\infty}^\infty \frac{\dd \kappa}{\sqrt{|\kappa|}}\, f^*(\kappa)\,g(\kappa)~. 
\end{equation} 
Notice that the kinetic energy and potential energy have anti-correlated signs, which is a crucial element in this calculation. The spectrum of principal-series eigenstates\footnote{See \cref{app:rep} for details. Here we focus on the even parity principal series.} satisfy
\beq
\ell_0^{\rm rel}\psi_n(\kappa)=-n\psi_n(\kappa)~,
\eeq
for $n\in\mathbb{Z}$, which has a general solution of the form 
\begin{equation}\label{eq:n2principalwavefunction}
\psi_n(\kappa)=\frac{|\kappa|^{\frac{3}{4}-\Delta}}{\pi\sqrt{2}}\int_{-\infty}^{\infty}\dd u\,(1+iu)^{-n-\Delta}(1-i u)^{n-\Delta}e^{-i \kappa u}~.
\end{equation}
It is relatively straightforward to check, using the substitution $u\rightarrow \tan\frac{v}{2}$ for $v\in (-\pi,\pi)$, that 
\beq
(\psi_n,\psi_m)=\delta_{nm}~,
\eeq
with respect to the inner product \eqref{eq:normn2kappa}. The above Fourier transform can be found in Volume I of \cite{bateman1954tables} and evaluates to
\begin{equation}
	\psi_n(\kappa)=\frac{2^{\frac{1}{2}-\Delta}}{|\kappa|^{\frac{1}{4}}}\times\begin{cases}-\frac{W_{-n,\frac{1}{2}-\Delta}\left(2\kappa\right)}{\Gamma(-n+\Delta)} &\kappa>0\\ \frac{W_{n,\frac{1}{2}-\Delta}\left(-2\kappa\right)}{\Gamma(n+\Delta)} &\kappa<0\end{cases}~,
\end{equation}
where $W_{a,b}(x)$ is the Whittaker function.
\paragraph{Summary:} We found that, to furnish the principal series for $N=2$, we needed to both analytically continue the Casimir eigenvalue $\Delta$, and the domain of the operators in \eqref{eq:conjugatedkappan2}. We will take these lessons and present the $N=3$ case now. What will follow will be very similar to the $N=2$ case, but with a slight twist.

\subsection{\texorpdfstring{$N=3$}{N=3}}
\subsubsection{Setup and original solution}

Let us now work out the example of $N=3$. We will (again) solve for the full spectrum of the problem, after which we will perform the analytic continuation in detail. This may seem redundant given that we solved for the entire spectrum at arbitrary $N$ in \cref{sec:Calobasics}. But working things out in detail here will help us illustrate analytic continuation to the principal series and how it mirrors the $N=2$ case; this will subsequently provide the roadmap for analytically continuing the problem at general $N$. 

\begin{figure}
    \centering
    \includegraphics[width=0.4\linewidth]{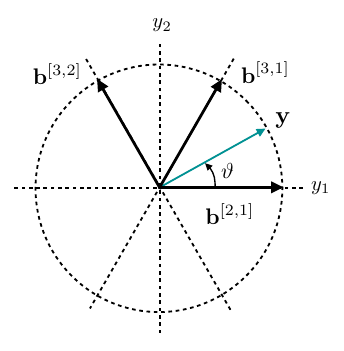}
    \caption{A depiction of the vectors $\mathbf{b}^{[i,j]}$ that enter in the potential of the $N=3$ Calogero model.}
    \label{fig:root_vectors}
\end{figure}

Once again, we ignore the center-of-mass degree of freedom and focus on the relative operators (c.f. \eqref{eq:conjugatedrel}):
\begin{equation}\label{eq:conjugatedreln3}
\mc{H}_{\rm rel}=-\frac{1}{2\,r}\partial_r\left( r\,\partial_r\right)+\frac{1}{2r^2}\hat{\mc{L}}^2_{S^{1}}~,\qquad \mc{K}_{\rm rel}=\frac{r^2}{2}~,\qquad \mc{D}_{\rm rel}=-\frac{i}{4}\left(\frac{1}{r}\partial_r\, r^{2}+r\,\partial_r\right)~,
\end{equation}
where
\beq 
	\hat{\mc{L}}^2_{S^{1}}
= -{\vdm^{-2\lambda}}\pa_\vartheta\left(\vdm^{2\lambda}\pa_\vartheta\right)+9\lambda^2~.
\eeq
These operators again form a closed $\slalg$ algebra with Casimir: 
\beq\label{eq:relcasimirn3}
	\mc{C}_2^{\rm rel}=\frac{1}{4}\left(\hat{\mc{L}}^2_{S^{1}}-1\right)~.
\eeq
To proceed, we need to compute the Vandermonde $\vdm$
\beq 
\vdm(\mathbf{y}) =\prod_{a}\left(\frac{\mathbf{b}^a\cdot \mathbf{y}}{r}\right)= \prod_a\,\cos\Theta^a~.
\eeq
The vectors $\mathbf{b}^a$, defined in \eqref{eq:ba}, are given by
\beq 
\mathbf{b}^{[2,1]} = \begin{pmatrix} 1 \\ 0\end{pmatrix}~, \qquad \mathbf{b}^{[3,1]} = \frac{1}{2}\begin{pmatrix} 1 \\ \sqrt{3}\end{pmatrix}~, \qquad \mathbf{b}^{[3,2]} = \frac{1}{2}\begin{pmatrix} -1 \\ \sqrt{3}\end{pmatrix}~,
\eeq
as displayed in fig. \ref{fig:root_vectors}. Note that these vectors correspond to (normalized) positive roots of the Lie algebra $\mathfrak{su}(3)$, and the angle between any two adjacent vectors is $\pi/3$. Taking 
\beq\mathbf{y}=r\begin{pmatrix} \cos\vartheta\\ \sin\vartheta\end{pmatrix}~,\eeq
implies
\beq
\cos\Theta^{[2,1]}=\cos(\vartheta)~,\qquad \cos\Theta^{[3,1]}=\cos\left(\frac{\pi}{3}-\vartheta\right)~,\qquad \cos\Theta^{[3,2]}=\cos\left(\frac{2\pi}{3}-\vartheta\right)~,
\eeq
and gives
\beq
\vdm(\mathbf{y})=-\frac{1}{4}\cos(3\vartheta) ~.
\eeq

Since the Casimir equation only depends on the angle $\vartheta$, the above system can be solved by separation of variables. Hence, we write our wavefunction as a product of radial and angular functions
\beq
\psi(r,\vartheta)=\phi(r)P(\vartheta)~.
\eeq
The first step to finding the spectrum of this model is to solve the Casimir equation, which translates to
\begin{equation}\label{eq:CasimirN3}
\mc{C}_2^{\rm rel}P(\vartheta)=\frac{1}{4}\left(-\frac{1}{\cos^{2\lambda}(3\vartheta)}\pa_\vartheta\left(\cos^{2\lambda}(3\vartheta)\pa_\vartheta\right)+9\lambda^2-1\right)P(\vartheta)=\Delta(\Delta-1)P(\vartheta)~.
\end{equation}
If we make the following clever choice for the $\vartheta$-dependence, 
\beq 
P(\vartheta)=P(\sin(3\vartheta))~,
\eeq
then the Casimir equation becomes:
\beq \label{eq:diffeqP}
\left(1-v^2\right)P''(v)-v(2\lambda+1)P'(v)+\frac{1}{9}\left(\left(1-2\Delta\right)^2-9\lambda^2\right)P(v) = 0~,
\eeq
in terms of the variable $v\equiv \sin(3\vartheta)$. Let us compare the above differential equation to the known Gegenbauer equation:
\begin{equation} \label{eq:defGegenb}
\left(1-v^2\right)P''(v)-v(2\alpha+1)P'(v)+m(m+2\alpha)P(v)=0~,
\end{equation}
whose solutions are the Gegenbauer polynomials $C_m^\alpha(v)$ for $m=0,1,2,\dots$ which are orthogonal with respect to to the following measure,
\begin{equation}
	\int_{-1}^1 \dd v \left(1-v^2\right)^{\alpha-\frac{1}{2}} C_n^\alpha(v)C_m^\alpha(v)=\frac{\pi\,2^{1-2\alpha}\Gamma(n+2\alpha)}{n!(n+\alpha)\Gamma(\alpha)^2}\delta_{nm}~.
\end{equation}
Comparing \eqref{eq:diffeqP} with \eqref{eq:defGegenb}, we see that the conformal Casimir equation admits orthonormal solutions for the following allowed values of $\Delta$:
\beq \label{eq:Deltam}
\Delta_m=\frac{1}{2}+\frac{3}{2}(m+\lambda)~,\qquad m=0,1,2,\dots
\eeq
and we note that the above expression for $m=0$ matches with $\Delta^\Omega_{\rm rel}$ in \eqref{eq:deltaomegarelanyN} for $N=3$. 
Thus, we can define a set of normalized harmonics,
\begin{equation} \label{eq:spheroidalharmonics}
 Y_m(\vartheta) \equiv 2^{3\lambda}\sqrt{\frac{3}{2\pi}\frac{m!(m+\lambda)\Gamma(\lambda)^2}{\Gamma(m+2\lambda)}}C_m^{\lambda}(\sin(3\vartheta))~,	
 \end{equation}   
 which are orthogonal with respect to the following measure,\footnote{Seemingly, the origin of the $\left(-\frac{\pi}{6},\frac{\pi}{6}\right)$ limits stems from keeping the angle $\vartheta+\pi/6$ within a chamber between the two root vectors $\mathbf{b}^{[2,1]}$ and $\mathbf{b}^{[3,1]}$. We believe that this is possibly due to the configuration space being redundant under permutations of the labels of the particles, so we need only consider this limited range of angles to get the whole configuration space.}
 \begin{equation} \label{eq:YYortho}
 	\int_{-\frac{\pi}{6}}^{\frac{\pi}{6}}\dd\vartheta\,|\vdm|^{2\lambda} \,Y_n(\vartheta)Y_m(\vartheta)=\delta_{nm}~.
 \end{equation}
Now that we have solved the angular problem, we just need to find the radial dependence. As usual we decompose our wavefunction into {harmonics} as follows, 
\begin{equation}
	\psi(r,\vartheta)=\sum_{m=0}^\infty b_m \phi_m(r) Y_m(\vartheta)~,
\end{equation}
where the $b_m$ are coefficients whose absolute squares sum to one, and $m$ labels the conformal family according to \eqref{eq:Deltam}. Restricting to a single value of $m$, the operator $\mc{H}_{\rm rel}$ in \eqref{eq:conjugatedreln3} acts as
\begin{equation}
\mc{H}_{\rm rel}=-\frac{1}{2\,r}\partial_r\left( r\,\partial_r\right)+\frac{4\Delta_m(\Delta_m-1)+1}{2r^2}~,
\end{equation}
so we are again dealing with a DFF model. Exactly as in the $N=2$ case, we can express the $\ell_0^{\rm rel}$ eigenstates at fixed $m$ in terms of Laguerre polynomials $L_k^\alpha(x)$. Let us define the functions
\begin{equation}
	\eta^m_k(r)\equiv \sqrt{\frac{2(k!)}{\Gamma(k+2\Delta_m)}}e^{-\frac{r^2}{2}}r^{2\Delta_m-1}L_k^{2\Delta_m-1}\left(r^2\right)~,
\end{equation}
for $k\in \mathbb{N}$ (note the extra factor of $r^{-1/2}$ compared to \eqref{eq:Laguerresolutions} due to a difference in kinetic terms), which are orthogonal with respect to the inner product
\begin{equation}
	\int_0^\infty \dd r \,r\, \eta^m_k(r)\eta^m_{k'}(r)=\delta_{kk'}~. 
\end{equation}
Defining the wavefunction:
\begin{equation}\label{eq:psimk}
\Psi_{mk}(r,\vartheta)=\eta^m_k(r)Y_m(\vartheta)~,
\end{equation}
it is easy to verify that
\begin{equation}
	\ell_0^{\rm rel}\,\Psi_{mk}(r,\vartheta)=(k+\Delta_m)\Psi_{mk}(r,\vartheta)~,
\end{equation}
and, in particular we have
\begin{equation}
\int_0^\infty\dd r\,r\int_{-\frac{\pi}{6}}^{\frac{\pi}{6}}\dd\vartheta\,|\vdm|^{2\lambda}\Psi_{mk}(r,\vartheta)\Psi_{m'k'}(r,\vartheta)=\delta_{mm'}\delta_{kk'}~.
\end{equation}

\subsubsection{Extra generators in \texorpdfstring{$N=3$}{N=3} problem}

We have managed to solve the $N=3$ problem without making use of the additional generators afforded to us, namely those originating in the fourth column in fig. \ref{fig:calogeroalgebra}. Since we have been ignoring the center-of-mass degree of freedom $X$ , we will drop all subscript `rel' in this section, but these are implied.

Let us focus on the third set of creation and annihilation operators. First, to make our notation more self contained, we define as in \eqref{eq:commutationKJ}:
\begin{equation}
[K,L_k]\equiv\frac{k}{2}M_{k-2}~,\qquad [K,M_k]\equiv\frac{k}{2}O_{k-2}~,
\end{equation}
then our creation and annihilation operators can be expressed simply as: 
\begin{equation}
a_3^\dagger=\frac{i}{4}\left(J_3+\frac{3}{2}L_1+\frac{3}{8}M_{-1}-\frac{1}{16}O_{-3}\right)~, \qquad a_3=-\frac{i}{4}\left(J_3-\frac{3}{2}L_1+\frac{3}{8}M_{-1}+\frac{1}{16}O_{-3}\right)~.
\end{equation}
Of course, we need the conjugated operators: 
\begin{equation}\label{eq:curlyA3}
\mathcal{A}_3^\dagger\equiv\vdm^{-\lambda}a_3^\dagger\vdm^{\lambda}~,\qquad \mathcal{A}_3\equiv\vdm^{-\lambda}a_3\vdm^{\lambda}~,
\end{equation}
whose action on the wavefunction $\Psi_{mk}$ defined in \eqref{eq:psimk} amounts to: 
\begin{align}
\mathcal{A}_3\Psi_{mk}&= A_{mk}\Psi_{(m-1)k}+B_{mk}\Psi_{(m+1)(k-3)}~,\nonumber\\ 
\mathcal{A}_3^\dagger\Psi_{mk}&= C_{mk}\Psi_{(m+1)k}+D_{mk}\Psi_{(m-1)(k+3)}~,
\end{align}
up to the unenlightening coefficients $(A_{mk},\dots,D_{mk})$. So as promised, these operators raise/lower the $\ell_0$ eigenvalue by $3/2$. Importantly, action by $\mathcal{A}_3$ and $\mathcal{A}_3^\dagger$ does not just raise or lower the conformal family, since the wavefunctions $\mathcal{A}_3\Psi_{mk}$ and $\mathcal{A}_3^\dagger\Psi_{mk}$ no longer diagonalize the Casimir operator $\mathcal{C}_2$.

\subsubsection{Analytic continuation to the principal series}
Unlike the case for $N=2$, the conformal families for the $N=3$ problem were determined as solutions to an eigenvalue problem. Recall that the conformal families in the previous section were found as solutions to \eqref{eq:CasimirN3}: 
\begin{equation}
	\frac{1}{4}\left[-\frac{1}{\cos^{2\lambda}(3\vartheta)}\partial_\vartheta\left(\cos^{2\lambda}(3\vartheta)\partial_\vartheta\right)+9\lambda^2-1\right] P(\vartheta)=\Delta(\Delta-1)P(\vartheta)~,
\end{equation}
and, as we explained there, the orthonormal solutions to the above differential equation are given in terms of a set of harmonics $P(\vartheta)=Y_m(\vartheta)$ defined in \eqref{eq:spheroidalharmonics} whose conformal dimensions were all in the discrete series: 
\beq
\Delta_m=\frac{1}{2}+\frac{3}{2}(m+\lambda)~, \qquad m\in\mathbb{Z}.
\eeq
The method we employed to define an $N=2$ theory in the principal series was to analytically continue $\lambda$. But we conclude now that this same trick will not work in the $N=3$ case, since we need to figure out how to analytically continue the above differential equation such that its spectrum is a set of eigenfunctions in the principal series.

After some thought, the solution turns out instead to be to analytically continue the angular variable $\vartheta$. Formally, if we make the change of variables $\vartheta\rightarrow i\chi-\pi/2$ and take
\begin{equation}\label{eq:deltasigma}
\Delta_\sigma=\frac{1}{2}(1+3i\sigma)~, \qquad \sigma \in \mathbb{R}~,
\end{equation}
as needed for the principal series, then the Casimir equation \eqref{eq:CasimirN3} becomes
\begin{equation} \label{eq:CasimireqN=3con}
	\frac{1}{4}\left[+\frac{1}{\sinh^{2\lambda}(3\chi)}\partial_\chi\left(\sinh^{2\lambda}(3\chi)\partial_\chi\right)+9\left(\lambda^2+\sigma^2\right)\right] P(\chi)=0~.
\end{equation}
This differential equation may not seem familiar, so let us massage it slightly. Let us first rewrite \eqref{eq:CasimireqN=3con} by taking $\chi\rightarrow\frac{2u}{3}$, which gives: 
\begin{equation}\label{eq:n3castau} P''(u)+2\lambda\left[\coth u+\tanh u\right]P'(u)+4(\lambda^2+\sigma^2) P(u)=0~.
\end{equation}
We can relate this differential equation to the more general theory of Jacobi functions which are defined as the set of functions that satisfy \cite{flensted1973convolution,Koornwinder1984JacobiFA,koornwinder1975two}:
\begin{equation}\label{eq:jacobi} P''(u)+\left[(2\alpha+1)\coth u+(2\beta+1)\tanh u\right]P'(u)+\left[(\alpha+\beta+1)^2+\gamma^2\right] P(u)=0~,
\end{equation}
for general $(\alpha,\beta,\gamma)$. The Jacobi functions constitute an interesting function space: they are basically a Fourier basis for the half-line. We therefore notice that \eqref{eq:n3castau} is a special case of \eqref{eq:jacobi} with $\alpha=\beta=\lambda-\tfrac{1}{2}$ and $\gamma=2\sigma$. The differential equation \eqref{eq:CasimireqN=3con} admits plane-wave normalizable solutions \cite{flensted1973convolution}:
\begin{equation}
	C_{i\sigma-\lambda}^{\lambda}(\cosh(3\chi))~,
\end{equation}
where $C_\nu^\gamma(x)$ are so-called Gegenbauer functions that are well-defined for complex values of the parameter $\nu\in \mathbb{C}$. These Gegenbauer functions can be expressed in terms of the more-familiar associated Legendre functions $P_\mu^\gamma(x)$ via the relation \cite{DLMF}
\begin{equation} 
C_{i\sigma-\lambda}^{\lambda}(\cosh(3\chi))=2^{\lambda-\frac{1}{2}}\frac{\Gamma\left(\frac{1}{2}+\lambda\right)\Gamma(\lambda+i\sigma)}{\Gamma\left(2\lambda\right)\Gamma(1-\lambda+i\sigma)}\sinh^{\frac{1}{2}-\lambda}(3\chi)P_{i\sigma-\frac{1}{2}}^{\frac{1}{2}-\lambda}(\cosh(3\chi))~.{}
\end{equation}
Remarkably, the orthogonality of these functions was studied in the 1950s by a geophysicist employed by a petroleum company in Texas, who discovered that \cite{van1954orthogonality} (see also \eqref{eq:orthogonalityP}): 
\begin{equation} \label{eq:orthoY}
	\int_1^\infty \dd u\, P_{i\sigma-\frac{1}{2}}^{\frac{1}{2}-\lambda}(u)P_{i\sigma'-\frac{1}{2}}^{\frac{1}{2}-\lambda}(u)=\left\lvert\frac{\Gamma(i\sigma)}{\Gamma(i\sigma+1-\lambda)}\right\rvert^2\left(\delta(\sigma-\sigma')+\delta(\sigma+\sigma')\right)~.
\end{equation}
Hence, if we define a set of wavefunctions
\begin{equation}\label{eq:Ychi}
Y_\sigma(\chi)\equiv2^{3\lambda}\sqrt{\frac{3}{2\pi}}\frac{\Gamma(\lambda)\Gamma(1-\lambda+i\sigma)^2}{\Gamma(i\sigma)\Gamma(\lambda+i\sigma)}C^\lambda_{i\sigma-\lambda}(\cosh(3\chi))~,
\end{equation}
then these are appropriately orthonormal with respect to the natural inner product:
\begin{equation}
\int_0^\infty \dd\chi\, |\vdm|^{2\lambda}\,Y^*_\sigma(\chi)Y_{\sigma'}(\chi)=\delta(\sigma-\sigma')~,
\end{equation}
where $|\vdm|=\frac{1}{4}\sinh(3\chi)$ is the natural analytically continued Vandermonde determinant and where we are now restricting $\sigma,\sigma'>0$.\footnote{The representations with negative $\sigma$ are isomorphic to positive $\sigma$ through the shadow map described in \cref{app:rep}.}

Now that we have solved the angular problem, what is left is to analytically continue the radial part, which will proceed exactly as before for $N=2$, namely we will perform the following analytic continuation:
\beq 
\frac{r^2}{2}\rightarrow\kappa~, \qquad \kappa\in(-\infty,\infty)~.
\eeq
Putting everything together, our task is to analyze the following $\slalg$-invariant quantum mechanics:
\begin{align}
\mc H_{\rm rel} & \equiv-\partial_\kappa \left(\kappa\,\partial_\kappa\right)+\frac{1}{4\kappa}\left[\frac{1}{\sinh^{2\lambda}(3\chi)}\partial_\chi\left(\sinh^{2\lambda}(3\chi)\partial_\chi\right)+9\lambda^2\right]~,\label{eq:hkappareln3}\\ \mathcal{K}_{\rm rel} & \equiv \kappa~,\\
\mathcal{D}_{\rm rel} & \equiv-\frac{i}{2}\left(\partial_\kappa\,\kappa +\kappa\,\partial_\kappa\right)~.
\end{align}
We will refer to the above system as the \emph{three-particle principal series Calogero model}.

We will present the spectrum of $N=3$ model in its entirety. Recall that we are looking for solutions to the equation:
\beq
\ell_0^{\rm rel}\Psi_{n\sigma}(\kappa,\chi)=-n\Psi_{n\sigma}(\kappa,\chi)~, \qquad \mc{C}_2^{\rm rel}\Psi_{n\sigma}(\kappa,\chi)=\Delta_\sigma(\Delta_\sigma-1)\Psi_{n\sigma}(\kappa,\chi)~,
\eeq
with $\Delta_\sigma$ defined in \eqref{eq:deltasigma} and $n\in \mathbb{Z}$. The solutions are relatively straightforward to write down. First we separate variables:
\begin{equation}
\Psi_{n\sigma}(\kappa,\chi)=\psi_n^\sigma(\kappa)Y_\sigma(\chi)~,
\end{equation}
where $Y_\sigma$ is as in \eqref{eq:Ychi}. The function $\psi_n^\sigma$ needs to satisfy
\beq
\frac{1}{2}\left[-\partial_\kappa \left(\kappa\,\partial_\kappa\right)-\frac{9\sigma^2}{4\kappa}+\kappa\right]\psi_n^\sigma(\kappa)=-n\psi_n(\kappa)~,
\eeq
and the solution is comparable with the $N=2$ case \eqref{eq:n2principalwavefunction}:
\begin{equation}
\psi_n^\sigma(\kappa)=\frac{|\kappa|^{\frac{1}{2}-\Delta_\sigma}}{\pi\sqrt{2}}\int_{-\infty}^{\infty}\dd u\,(1+iu)^{-n-\Delta_\sigma}(1-i u)^{n-\Delta_\sigma}e^{-i \kappa u}~.
\end{equation}
Volume I of \cite{bateman1954tables} allows us to write the integral in terms of Whittaker functions:
\begin{equation}
	\psi_n^\sigma(\kappa)=\frac{2^{\frac{1}{2}-\Delta_\sigma}}{\sqrt{|\kappa|}}\times\begin{cases}-\frac{W_{-n,\frac{1}{2}-\Delta_\sigma}\left(2\kappa\right)}{\Gamma(-n+\Delta_\sigma)} &\kappa>0\\ \frac{W_{n,\frac{1}{2}-\Delta_\sigma}\left(-2\kappa\right)}{\Gamma(n+\Delta_\sigma)} &\kappa<0\end{cases}~.
\end{equation}
By construction, we know that the wavefunctions are normalized as follows: 
\begin{equation}
\int_{-\infty}^\infty\dd \kappa\int_0^\infty\dd\chi\,|\vdm|^{2\lambda}\,\Psi_{n\sigma}^*(\kappa,\chi)\Psi_{m\sigma'}(\kappa,\chi)=\delta_{nm}\,\delta(\sigma-\sigma')~,
\end{equation}
where we remind the reader that $|\vdm|=\frac{1}{4}\sinh(3\chi)$~. The upshot of this analysis is that we have found a basis of normalizable wavefunctions which solve the conformal quantum mechanics defined in \eqref{eq:hkappareln3} and carrying the spectrum of the principal series representation.

Instead of working in a basis of $\ell_0^{\rm rel}$ eigenstates, we could equally work in a basis of $\mathcal{H}_{\rm rel}$ eigenstates (see Appendix A of \cite{Anous:2020nxu} for what these wavefunctions look like). This would allow us to compute the Harish-Chandra character for this $\slalg$-invariant quantum mechanics, as in \cite{Anninos:2020hfj,Comtet:1984mm,Comtet:1986ki}, but we leave this to future work.

\subsubsection{Analytic continuation of \texorpdfstring{$\mathcal{J}_3^{\rm rel}$}{J3rel}?}\label{subsubsec:J3rel}
One point emphasized in \cref{sec:exact_quantum_spectrum_algebraic_method} was that the entire spectrum of the Calogero model follows from the organization of its current algebra. Above, we did not make use of the extra currents in solving for the spectrum of the analytically continued $N=3$ model. 
Here, for the sake of completeness, we investigate the analytic continuation of the extra current for the $N=3$ model. Let us write down the conjugated relative current $\mathcal{J}_3^{\rm rel}$ in the new variables:
\begin{align}
\mathcal{J}_3^{\rm rel}=\frac{i}{2}\sqrt{\frac{3}{\kappa}}&\left\lbrace\frac{\kappa}{\sinh^{\lambda-1}(3\kappa)}\partial_\kappa^2\partial_\chi\left(\sinh^{\lambda}(3\chi)\right)\right.\nonumber\\
&~~~~-\frac{1}{2}\partial_\kappa\left[\frac{\cosh^{1/3}(3\chi)}{\sinh^{2\lambda}(3\chi)}\partial_\chi\left(\cosh^{2/3}(3\chi)\sinh^{2\lambda}(3\chi)\right)+3\lambda(3\lambda+2)\cosh(3\chi)\right]\nonumber\\
&~~~~+\frac{\cosh(3\chi)}{\kappa}\left[-\frac{2}{3}\kappa^3\partial^3_\kappa+\frac{3\lambda+2}{4\cosh^{\frac{3\lambda+4}{9}}(3\chi)}\partial_\chi\left(\cosh^{\frac{3\lambda+4}{9}}(3\chi)\partial_\chi\right)\right.\nonumber\\ 
&\qquad\qquad\qquad~~~~\left.\left.+\frac{3\lambda(\lambda+1)}{2\sinh(3\chi)}\partial_\chi\left(\cosh(3\chi)\right)+\frac{\tanh(3\chi)}{12}\partial_\chi^3+\frac{\lambda}{4}\left(9\lambda^2-10\right)\right]\right\rbrace ~,
\end{align}
and we have checked that this indeed commutes with \eqref{eq:hkappareln3}, as expected. However, it seems that this operator is not Hermitian on the domain $\kappa \in (-\infty,\infty)$, and therefore we do not expect to be able to use it to generate the spectrum, as we did in the discrete series. The reason for this is the pesky $\kappa^{-1/2}$ sitting outside of the curly braces, which crosses a branch cut when $\kappa$ becomes negative. We believe that this happens for all the higher currents under this analytic continuation, although we have not presented a general proof. 

{An alternative interpretation could be that the loss of Hermiticity of $\mathcal{J}_3^{\rm rel}$ may indicate that we should think of it more akin to a creation or annihilation operator (similar to $\mathcal{A}_3$ and $\mathcal{A}_3^\dagger$ in \eqref{eq:curlyA3}). We have not explored this possibility, although it is a tantalizing one for maintaining the integrability of the model.}

Regardless of {whether integrability is preserved or lost in the continuation to the principal series}, we will now explain how our solution for the three-particle principal series Calogero model sets a road map for defining the Calogero model in the principal series for general $N$.

\subsection{The principal series Calogero model for general \texorpdfstring{$N$}{N} and \texorpdfstring{$\lambda$}{lambda}}

We now have all the ingredients ready to present the analytic continuations required to define the principal series Calogero model for general $N$ and $\lambda$. To begin, recall that in the original model we had introduced spherical coordinates on the $\mathbf{y}$-vector in $\mathbb{R}^{N-1}$, which involved the use of spherical coordinates on the $(N-2)$-sphere via the recursive formula
\beq \label{eq:omega_i}
\omega_1 = \cos \vartheta~, \qquad \omega_i = \nu_i \sin \vartheta~, \qquad i = 2,\ldots, N-1~,
\eeq
where the $\nu_i$ represent coordinates on the $(N-3)$-sphere (see equation \eqref{eq:nm2sphere}). Hence we have defined things  such that:  \begin{equation}\sum_{i=1}^{N-1}(\omega_i)^2=\sum_{i=2}^{N-1}(\nu_i)^2=1~.\end{equation} 
Moreover, recall that the conformal families of the original model were found as solutions to 
\beq
\mc{C}_2^{\rm rel}P(\Omega)\equiv\frac{1}{4}\left(\hat{\mc{L}}^2_{S^{N-2}}+\frac{(N-5)(N-1)}{4}\right)P(\Omega)=\Delta(\Delta-1)P(\Omega)~,
\eeq
where the angular operator $\hat{\mc{L}}^2_{S^{N-2}}$ 
\beq 
	\hat{\mc{L}}^2_{S^{N-2}}
\equiv -\frac{\vdm^{-2\lambda}}{\sqrt{g}}\pa_\mu\left(\vdm^{2\lambda}\sqrt{g}g^{\mu\nu}\pa_\nu\right)+\lambda\frac{N(N-1)}{2}\left(\lambda\frac{N(N-1)}{2}+N-3\right)~,
\eeq
was defined in \eqref{eq:reducedangH}~ and where $g_{\mu\nu}$ is the metric on the unit $S^{N-2}$. The above Casimir operator only admits normalizable solutions in the discrete series. Below we will detail the analytic continuation needed such that $\mc{C}_2^{\rm rel}$ admits solutions in the principal series. 

For the moment, let us suppose that we have already managed to find the appropriate analytic continuation for $\hat{\mc{L}}^2_{S^{N-2}}$. Then, on any subsector of fixed conformal dimension $\Delta_\gamma=\frac{1}{2}(1+i\gamma)$, in the principal series we can work with \eqref{eq:conjugatedrel} and take the, by now, familiar analytic continuation
\beq 
\frac{r^2}{2}\rightarrow\kappa~, \qquad \kappa\in(-\infty,\infty)~,
\eeq
which leads to: 
\begin{align}
\mc H_{\rm rel} & \equiv-\kappa\partial_\kappa^2 -\frac{N-1}{2}\partial_\kappa+\frac{1}{\kappa}\left(\Delta_\gamma+\frac{N-5}{4}\right)\left(\Delta_\gamma-\frac{N-1}{4}\right)~,\\
\mc{K}_{\rm rel} & \equiv \kappa~,\\
\mathcal{D}_{\rm rel} & \equiv-i\left(\kappa\,\partial_\kappa+\frac{N-1}{4}\right)~.
\end{align}
These operators are Hermitian with respect to the inner product: 
\beq
(f,g)=\int_{-\infty}^{\infty}\dd \kappa\,|\kappa|^{\frac{N-3}{2}}f^*(\kappa)\, g(\kappa)~,
\eeq
and the eigenfunctions, satisfying $\ell_0^{\rm rel}\psi_n^\gamma(\kappa)=-n\,\psi_n^\gamma(\kappa)$, are given by
\begin{equation}\label{eq:PSkappaWF1}
\psi_n^\gamma(\kappa)=\frac{|\kappa|^{\frac{5-N}{4}-\Delta_\gamma}}{\pi\sqrt{2}}\int_{-\infty}^{\infty}\dd u\,(1+iu)^{-n-\Delta_\gamma}(1-i u)^{n-\Delta_\gamma}e^{-i \kappa u}~,
\end{equation}
and can equivalently be written as: 
\begin{equation}\label{eq:PSkappaWF}
\psi_n^\gamma(\kappa)=\frac{2^{\frac{1}{2}-\Delta_{\gamma}}}{|\kappa|^{\frac{N-1}{4}}}\times\begin{cases}-\frac{W_{-n,\frac{1}{2}-\Delta_\gamma}\left(2\kappa\right)}{\Gamma(-n+\Delta_\gamma)} &\kappa>0\\ \frac{W_{n,\frac{1}{2}-\Delta_\gamma}\left(-2\kappa\right)}{\Gamma(n+\Delta_\gamma)} &\kappa<0\end{cases}~.
\end{equation}
So with the radial part of the equations  out of the way, what remains to complete our story is to detail the correct analytic continuation of $\hat{\mc{L}}^2_{S^{N-2}}$ .

\subsubsection{Free limit of the principal series Calogero model}\label{subsubsec:free}
To make the general case more accessible, we will start by analyzing the limit $\lambda\rightarrow0$ in which the Calogero model reduces to a theory of $N$ free particles on a line. In this limit, the angular operator appearing in $\mc{C}_2^{\rm rel}$ reduces to the Laplacian on the $(N-2)$-sphere
\beq \label{eq:spherelaplacian}
\lim_{\lambda\to0}\hat{\mc{L}}^2_{S^{N-2}}= -\nabla^2_{S^{N-2}}=-\frac{1}{\sin^{N-3}(\vartheta)}\, \partial_{\vartheta}\left(\sin^{N-3}(\vartheta )\,\partial_{\vartheta} \right)+\frac{1}{\sin^2\vartheta}\nabla^2_{S^{N-3}}~,
\eeq
where $\nabla^2_{S^{N-3}}$ is the Laplace-Beltrami operator on the $(N-3)$-sphere.
Hence, the Casimir equation for the non-interacting model becomes:
\beq \label{eq:sphericalCasimir}
\frac{1}{4}\left[-\nabla^2_{S^{N-2}}+\frac{(N-5)(N-1)}{4}\right]P(\omega) = \Delta(\Delta-1)P(\omega)~.
\eeq
whose solutions are the spherical harmonics on the $N-2$ sphere $Y_{L\vec{m}}(\Omega)$ which satisfy: 
\beq
-\nabla^2_{S^{N-2}}Y_{L\vec{m}}(\omega)=L(L+N-3)Y_{L\vec{m}}(\omega)~.
\eeq
Simple algebra reveals that $\Delta=\frac{L}{2}+\frac{N-1}{4}$ for these solutions, and the $L=0$  state matches the $\lambda\to0$ limit of $\Delta^\Omega_{\rm rel}$ of \eqref{eq:deltaomegarelanyN}, as expected. 

We would now like to convert the left-hand side of the above Casimir equation to an operator whose solutions lie in the principal series representation. 
We will do this by taking $\vartheta\rightarrow iu$ similar to what we did in the $N=3$ case. 
With this replacement, the natural embedding coordinates \eqref{eq:omega_i} become: 
\beq
\hat\omega_1=\cosh u~,\qquad \hat\omega_i = \nu_i \sinh u~, \qquad i = 2,\ldots, N-1~,
\eeq
with the $\nu_i$ unchanged, but where these coordinates $\hat\omega_i$ instead satisfy: 
\beq
\left(\hat\omega_1\right)^2-\sum_{i=2}^{N-1}\left(\hat\omega_i\right)^2=1~,
\eeq
meaning these coordinates
parametrize the upper sheet of a two-sheeted hyperboloid with the variable $u\in[0,\infty)$  corresponding to the non-compact direction. The metric on this hyperboloid is:
\beq\label{eq:hyperbolicmetric}
\dd s^2\equiv G_{\mu\nu}\dd x^\mu\dd x^\nu = \dd u^2+\sinh^2(u) \dd\Omega_{N-3}^2~,
\eeq
and the corresponding Laplace-Beltrami operator $\nabla^2_{\mathbb{H}^{N-2}}$, which is given by 
\beq \label{eq:hyperbolicLaplacian}
\nabla^2_{\mathbb{H}^{N-2}}\equiv\frac{1}{\sqrt{G}}\partial_\mu\left(\sqrt{G}G^{\mu\nu}\partial_\nu\right)= \frac{1}{\sinh^{N-3}(u)}\, \partial_{u}\left(\sinh^{N-3}(u)\, \partial_{u} \right)+\frac{1}{\sinh^2 u}\nabla^2_{S^{N-3}}~.
\eeq
Under the analytic continuation $\vartheta\to iu$ the Laplace-Beltrami operator \eqref{eq:spherelaplacian} becomes
\beq
-\nabla^2_{S^{N-2}}\rightarrow+\nabla^2_{\mathbb{H}^{N-2}}~,
\eeq
and hence, the corresponding Casimir equation is now 
\beq \label{eq:hyperbolicCasimir}
\frac{1}{4}\left[\nabla^2_{\mathbb{H}^{N-2}}+\frac{(N-5)(N-1)}{4}\right]P(\hat\omega) = \Delta(\Delta-1)P(\hat\omega)~.
\eeq
Following \cite{perelomov1977generalized}, we will now argue that this equation admits solutions in the principal series representation. This can be done by making the following ansatz:
\beq 
P(\hat\omega) = \sinh^\ell(u) Y_{\ell\vec{m}}(\nu)\Phi(u)~,
\eeq
where $Y_{\ell\vec{m}}(\nu)$  are the spherical harmonics  on the $(N-3)$-sphere, satisfying the  eigenvalue equation 
\beq 
-\nabla^2_{S^{N-3}}Y_{\ell\vec m}= \ell(\ell+N-4) Y_{\ell\vec m}~.
\eeq
 Plugging this into \eqref{eq:hyperbolicCasimir} and setting $\Delta=\frac{1}{2}(1+i\gamma)$ we find $\Phi(u)$ must satisfy:
\beq \label{eq:Philambda}
\Phi''(u)+\left(2\ell +{N-3}\right)\coth u \, \Phi'(u)+\left(\left(\ell+\frac{N-3}{2}\right)^2+\gamma^2\right)\Phi(u)=0~.
\eeq
Let us compare this with the defining equation for the Jacobi functions that was presented in \eqref{eq:jacobi}, which we repeat here for convenience:
\begin{equation}
P''(u)+\left[(2\alpha+1)\coth u+(2\beta+1)\tanh u\right]P'(u)+\left[(\alpha+\beta+1)^2+\gamma^2\right] P(u)=0~.
\end{equation}
From this we determine that \eqref{eq:Philambda} is a Jacobi differential equation with $\alpha=\ell+\frac{N-4}{2}$, $\beta=-\frac{1}{2}$~ and, following \cite{Koornwinder1984JacobiFA,van1954orthogonality,koornwinder1975two,perelomov1977generalized}  we therefore know that there exists solutions which are orthonormal with respect to the appropriate measure, just as in the interacting $N=3$ case.

\subsubsection{Adding interactions}
To add the interaction we simply need to provide an analytically continued definition of the Vandermonde determinant. We will do this as follows. First define the following vector in $\mathbb{C}^{N-1}$: 
\beq
\hat{\mathbf{z}}\equiv\left(\cosh u, i\,\vec{\nu}\,\sinh u \right)~,
\eeq
where recall $\vec{\nu}$ is an $(N-2)$-vector denoting a position on the unit $S^{N-3}$. We define the appropriate Vandermonde determinant for the principal series as follows: 
\beq
\widehat{\vdm}\equiv \prod_{a}\mathbf{b}^a\cdot \hat{\mathbf{z}}~,
\eeq
where $\mathbf{b}^a$ were the set of $N(N-1)/2$ vectors defined in \eqref{eq:ba} and the dot denotes the standard Euclidean product of these vectors. This ensures that $\widehat{\vdm}$ can be obtained as the analytic continuation in $\vartheta\rightarrow i u$ of the standard Vandermonde determinant defined in \eqref{eq:vanddef}. Defined in this way, $\widehat{\vdm}$ is either pure real or pure imaginary, which is a consequence of the identity $\Theta^{[2,1]}=\vartheta$ prior to analytic continuation. In detail, using the properties of the root vectors $\mathbf{b}^a$,\footnote{Namely,
\beq
    \mathbf{b}^a\cdot \hat{\mathbf z}=b_1^a\cosh u+i\sinh u\,\zeta^a~,
\eeq
with
\beq
    \mathbf{b}^{[2,1]}=(1,0,\ldots,0)~,\qquad b^{[i>2,1]}_1=-b^{[i>2,2]}_1=\frac12~,\qquad b^{[i>2,j>2]}_1=0~,\qquad \zeta^{[i,1]}=\zeta^{[i,2]}~.
\eeq
}
\begin{multline}\label{eq:ACvan}
    \widehat{\vdm}=(-1)^{N-2}i^{\frac{(N-3)(N-2)}{2}}\cosh u(\sinh u)^{\frac{(N-3)(N-2)}{2}}\left[\prod_{i>j>2}\zeta^{[i,j]}\right]\\\times\left[\prod_{j>2}\left(\frac{1}{4}\cosh^2u+\zeta^{[j,1]2}\sinh^2u\right)\right]~,
\end{multline}
where
\beq
    \zeta^{[i,j]}\equiv \sum_{k=2}^{N-1}\mathbf{b}^{[i,j]}_{k}\nu^k~.
\eeq
With this in hand, let us define the appropriate principal series generalization of \eqref{eq:reducedangH}: 
\beq\label{eq:reducehypang}
	\hat{\mc{L}}^2_{\mathbb{H}^{N-2}}
\equiv +\frac{\widehat{\vdm}^{-2\lambda}}{\sqrt{G}}\pa_\mu\left(\widehat{\vdm}^{2\lambda}\sqrt{G}G^{\mu\nu}\pa_\nu\right)+\lambda\frac{N(N-1)}{2}\left(\lambda\frac{N(N-1)}{2}+N-3\right)~,
\eeq
where we remind the reader that $G_{\mu\nu}$ is the metric on the upper sheet of the two-sheeted $(N-2)$-dimensional hyperboloid defined in \eqref{eq:hyperbolicmetric}, and we clearly see that the phase in $\widehat\vdm$ cancels between the numerator and denominator of \eqref{eq:reducehypang}. With this we  can finally write down the full \emph{$N$-particle principal series Calogero model:} 
\begin{align}\label{eq:NPSCalofull}
\mc H & \equiv-\kappa\partial_\kappa^2 -\frac{N-1}{2}\partial_\kappa+\frac{1}{4\kappa}\hat{\mc{L}}^2_{\mathbb{H}^{N-2}}~,\\
\mc{K} & \equiv \kappa~,\\
\mathcal{D} & \equiv-i\left(\kappa\,\partial_\kappa+\frac{N-1}{4}\right)~.
\end{align}
These operators are Hermitian with respect to the following inner product: 
\beq\label{eq:NPSinnerprod}
(f,g)=\int_{-\infty}^{\infty}\dd \kappa\,|\kappa|^{\frac{N-3}{2}}\int_0^\infty\dd u\int \dd \Omega_{N-3} \sqrt{G}|\widehat\vdm|^{2\lambda}f^*(\kappa,u,\vec\nu)\, g(\kappa, u,\vec\nu)~,
\eeq
and furnish the $\slalg$ algebra \eqref{eq:calogerosl2} with Casimir operator
\beq
	\mc{C}_2\equiv \frac{1}{2}\left(\mc{H}\,\mc{K}+\mc{K}\,\mc{H}\right)-\mc{D}^2=\frac{1}{4}\left(\hat{\mc{L}}^2_{\mathbb{H}^{N-2}}+\frac{(N-5)(N-1)}{4}\right)~.
\eeq
The $\ell_0^\text{rel}$ eigenfunctions are given by
\beq
    \Psi^{\gamma}_n(\kappa,u,\vec\nu)=\psi_n^\gamma(\kappa)P_\gamma(u,\vec\nu)~,
\eeq
with $\psi_n^\gamma(\kappa)$ given in \eqref{eq:PSkappaWF} and $P_\gamma(u,\vec\nu)$ satisfying the Casimir equation
\beq\label{eq:NPSCasEq1}
    \mathcal C_2^\text{rel}P_\gamma(u,\vec\nu)\equiv\frac{1}{4}\left(\hat{\mathcal L}^2_{\mathbb H^{N-2}}+\frac{(N-5)(N-1)}{4}\right)P_\gamma(u,\vec \nu)=\Delta_{\gamma}(\Delta_{\gamma}-1)P_\gamma(u,\vec\nu)~,
\eeq
or equivalently
\beq\label{eq:NPSCasEq2}
    \left[\frac{\widehat{\vdm}^{-2\lambda}}{\sqrt{G}}\pa_\mu\left(\widehat{\vdm}^{2\lambda}\sqrt{G}G^{\mu\nu}\pa_\nu\right)+\left(\lambda\frac{N(N-1)}{2}+\frac{N-3}{2}\right)^2+\gamma^2\right]P_\gamma(u,\vec\nu)=0~,
\eeq
with $\widehat{\vdm}$ given explicitly by \eqref{eq:ACvan}. While we have not managed to solve the Casimir equation \eqref{eq:NPSCasEq2} for general $N$ and $\lambda$, by construction $\Psi_n^\gamma$ satisfies
\beq
    \ell_0^\text{rel}\Psi^\gamma_n(\kappa,u,\vec\nu)=-n\,\Psi^\gamma_n(\kappa,u,\vec\nu)~.
\eeq
We can furthermore demonstrate that the wavefunctions are plane-wave normalizable with respect to the inner product \eqref{eq:NPSinnerprod}:
\beq
(\Psi_n^{\gamma_1},\Psi_m^{\gamma_2})=\delta_{mn}\delta(\gamma_1-\gamma_2)~,
\eeq
where we have taken $\gamma_{1,2}>0$.\footnote{Again, representations with negative $\gamma$ are isomorphic to those with positive $\gamma$ through the shadow isomorphism discussed in \cref{app:rep}.} This follows from the defining differential equation, which we demonstrate in appendix \ref{app:PSWFsortho}. The upshot of this section is that for general $N$, we have defined an interacting conformal quantum mechanics that admits (plane-wave) normalizable wavefunctions carrying the quantum numbers of the principal series representation.

\section{Discussion}\label{sec:disc}

In this paper we explored a system of interacting, multi-particle, conformal quantum mechanics with a spectrum furnished by a continuum of principal series representations of $\slalg$. Our line of investigation consisted of analytically continuing the Calogero model, an integrable model of interacting quantum mechanics of particles arranged on a line. Along the way we have collected and reviewed standard features of the Calogero model, namely the structure of the symmetry generators responsible for its integrability, its solution, and its partition function. Central to this solution was splitting the center-of-mass and relative dynamics. We then further explained how the relative dynamics of the model could be analytically continued to display the principal series spectrum. We worked this continuation and its solution explicitly for the $N=2$ and $N=3$ models while providing a concrete roadmap for the solution (including its spectrum and inner product) for general $N$ model. Let us now discuss some open questions that follow naturally from this work.

\subsection*{Explicit states and integrability?}

While we have solved the principal series Calogero model explicitly for $N=2$ and $N=3$, we have left the solution for generic $N$ implicit upon solving the hyperbolic Casimir equation, \eqref{eq:NPSCasEq2}. Of course the spectrum of $\mathcal C_2^\text{rel}$ (and thus that of $\hat{\mathcal L}^2_{\mathbb H^{N-2}}$ and the Hamiltonian) are fixed by the supposition that our solutions furnish the principal series representation. We have also shown, based on general features of the Casimir equation, that such solutions are plane-wave normalizable. Regardless, one might speculate if solutions can be constructed explicitly. In our review of the standard Calogero model in \cref{sec:Calobasics}, the particular details of the eigenfuctions have been made immaterial: instead, we have utilized the $N$ independent symmetry currents to completely generate the spectrum of the model and as such, states can be constructed explicitly as \cref{eq:CalogeroFock}. For our principal series continuation, we do not have such a luxury. As we demonstrated in \cref{subsubsec:J3rel}, the continuation of the first of the higher currents fails to be Hermitian on the domain $\kappa\in(-\infty,\infty)$, displaying branch cuts for negative $\kappa$. Thus it is not obvious that we can utilize it to generate positive norm states. This is perhaps not surprising as the spectrum of the principal series model is continuous and we do not expect it to be generated by a discrete set of generators. Regardless, in principle we still retain a large set of commuting (albeit non-Hermitian) operators at general $N$ and whether this fact has leverage for constructing solutions remains a question worth investigating.

\subsection*{Partition functions}

In \cref{eq:partitionfunccalogero} we presented the partition function of the standard $N$-particle Calogero model. Although this partition function only explicitly depends on $\lambda$ through the ground state energy, $\Delta_\Omega$, given in \cref{eq:DeltaOmegaboxed}, it is discontinuous from the non-interacting partition function which is given by the product of decoupled harmonic oscillators. We were able to deduce this partition function, again, by directly utilizing the extra generators built from the symmetry algebra.

For the principal series model we have lost access to integrability but we still know that its spectrum is fixed by the $\slalg$ representation theory. For principal series representations, the appropriate notion of a representation trace is distributional and defined by a Harish-Chandra character (as  popularized in \cite{Anninos:2020hfj}), which is most naturally taken in the basis diagonalizing $\mc H$ (as opposed to $\ell_0^\text{rel}$). While we have not written this basis or computed the corresponding Harish-Chandra characters in this paper, this should be a technically straightforward task. The computation of the partition function of this model however involves two more subtle issues. The first is that the solution space of \eqref{eq:NPSCalofull} is generated by a continuum of principal series representations. Thus the partition function should involve an integral over Harish-Chandra characters of distinct principal series representations with an appropriate weight. The second is that there may be a degeneracy of solutions yielding the same Casimir equation \cref{eq:NPSCasEq1}. This was indeed the case for the non-interacting limit of our model in \cref{subsubsec:free}: we found a solution to the Casimir equation for every spherical harmonic on the $(N-3)$-sphere. However as we have alluded to just above, one lesson from the standard Calogero model is that the state space of the non-interacting theory may not smoothly connect to the that of the interacting theory and so we cannot yet draw any conclusions about the degeneracy of solutions to \cref{eq:NPSCasEq1} from its non-interacting limit. Having a more explicit route of constructing solutions (in line with our previous discussion point) would aid towards addressing this question. In general constructing the partition function of our model and explicating its dependence on both $\lambda$ and $N$ is an important open question we plan to revisit.

\subsection*{The large $N$ limit}

Closely related to the above discussion point is what arises in the large $N$ limit of our analytically continued model. We recall that the partition function of the standard Calogero model displays features of a local quantum field theory in the $N\rightarrow\infty$ limit, that of a 1+1-$d$ $U(1)$ symmetric chiral CFT. In this case the $N\rightarrow\infty$ is akin to a continuum limit of the target space of $N$-particle quantum mechanics.

It is interesting to speculate if the $N\rightarrow\infty$ limit of the principal series Calogero model retains any realization as a local quantum field theory. Such a realization would be a potentially interacting and solvable $\slalg$ symmetric field theory with states living in the principal series. This is an intriguing prospect made more intriguing by connections to two-dimensional de Sitter physics, a motivation we described in the introduction to this paper. One obvious obstacle to investigating the large $N$ limit is first constructing the partition function of the interacting principal series model, as we discussed in the previous point. Additional to this is the obstacle of reincorporating the center of mass degree of freedom. For the standard Calogero model, the center of mass is essential part of both realizing the target space of the relative system quantum mechanics as particles on a line, as well as recognizing the $N\rightarrow\infty$ partition function as a chiral CFT partition function. We expect that in the principal series model, the center of mass to a play a similar role in both contexts. In this case the target space is more subtle as we have analytically continued the relative angles between particles. It is possible that realizing both a sensible target space and a sensible large $N$ partition function might require analytically continuing the center of mass as well, although at this point we leave this as a speculation. 

\section*{Acknowledgements}
We are grateful to Dionysios Anninos, Frederik Denef,  Diego Hofman, Austin Joyce, Joe Minahan, Gui Pimentel, Zimo Sun, Erik Verlinde, and Stathis Vitouladitis for discussions and comments. We especially thank R. Esp\'indola and B. Najian for collaboration during the initial stages of this project.  
TA is supported by UKRI Future Leaders Fellowship ``The materials approach to quantum spacetime'' under reference MR/X034453/1. JRF is partially supported by STFC consolidated grants ST/T000694/1 and ST/X000664/1, and partially by Simons Foundation Award number 620869. JvdH acknowledges support from the National Science and Engineering Research Council of Canada (NSERC) and the Simons Foundation via a Simons Investigator
Award.

\appendix 

\section{Review of \texorpdfstring{$\slalg$}{sl2r} representation theory}
\label{app:rep}

In this section, we review the unitary representation theory of $\slalg$. An ultimate goal of this section is to introduce the basic features of the principal series representations and distinguish them from complementary and discrete series representations.

The $\slalg$ algebra is given by Hermitian generators $\big\{H,D,K\big\}$ satisfying the following commutation relations.
\beq \label{eq:sl2comms}
	[D,H]=i H~,\qquad [D,K]=-i K~,\qquad [K,H]=2i D~.
\eeq
This is a real basis of $\slalg$ and we will construct representations with inner products satisfying
\beq\label{eq:HKDherm}
    H^\dagger=H~,\qquad K^\dagger=K~,\qquad D^\dagger=D~.
\eeq
Central to the classification of such representations is the quadratic Casimir, 
\beq\label{eq:sl2Cas}
    C_2\equiv \frac12\left(HK+KH\right)-D^2~,
\eeq
which commutes with all generators and so is a constant on irreducible representations. We can construct a representation in the standard manner by treating $D$ as a Cartan element and looking for eigenstates that are highest-weight. In this context, such states are called {\it primary} and satisfy
\beq\label{eq:primarystatedef}
    D|\Delta,0\rangle=i\Delta|\Delta,0\rangle~,\qquad K|\Delta,0\rangle=0~,
\eeq
where $\Delta$ is known as the {\it conformal dimension}. Acting on primary states 
\beq
    C_2|\Delta,0\rangle=\Delta(\Delta-1)|\Delta,0\rangle~,
\eeq
which labels the representation. From this point, we will omit the label ``$\Delta$'' from the state when it is clear which representation it is a part of. Reality of the Casimir implies two possible cases for the conformal dimension, $\Delta\in\mathbb R$, or $\Delta=\frac{1}{2}+i\nu$ with $\nu\in\mathbb R$.

A useful strategy \cite{Sengor:2019mbz,Sun:2021thf} for building representations for either case of $\Delta$ is to construct a set of ladder operators of the form
\beq
    l_0\equiv \frac12\left(H+K\right)~,\qquad l_\pm\equiv\frac12\left(H-K\right)\mp iD~,
\eeq
satisfying
\beq
    [l_+,l_-]=2l_0~,\qquad [l_\pm,l_0]=\pm l_\pm~.
\eeq
Thus $l_+$ lowers the $l_0$ eigenvalue while $l_-$ raises it. Note that \eqref{eq:HKDherm} implies the reality conditions 
\beq\label{eq:Lherm}
    l_0^\dagger =l_0~,\qquad l_\pm^\dagger=l_\mp~.
\eeq
Furthermore, under exponentiation $l_0$ generates the maximal compact subgroup, $SO(2)\subset SL(2,\mathbb R)$, and so its eigenvalues are either integer or half-integer valued (which we will call even or odd parity, respectively). We can move to the eigenbasis of $l_0$, $\{|r\rangle\}_{r\in\mathbb Z~\text{or}~ r+\frac12\in\mathbb Z}$ satisfying
\beq\label{eq:Laction}
    l_0|r\rangle=-r|r\rangle~,\qquad l_\pm|r\rangle=-\left(r\pm \Delta\right)|r\pm 1\rangle~.
\eeq
For completeness, we record the relation between the $\{|r\rangle\}$ basis and the primary state, \eqref{eq:primarystatedef}, as \cite{Anous:2020nxu}
\beq
    |r\rangle=\int_{-\infty}^\infty\dd t\,(1+it)^{-r-\Delta}(1-it)^{r-\Delta}e^{-iHt}|0\rangle~.
\eeq
It is easy to see that for an inner product satisfying \eqref{eq:Lherm}, the relation $\langle r|l_-|r+1\rangle=\langle r+1|l_+|r\rangle^\ast$ implies
\beq\label{eq:normratio}
    \frac{\langle r+1|r+1\rangle}{\langle r|r\rangle}=\frac{r+1-\Delta}{r+\Delta^\ast}~.
\eeq
We can build a positive inner product for all states of the representation through induction if and only if this ratio is positive for all $r$. 

When $\Delta=\frac12+i\nu$ this ratio is one for all $r$ and so the representation is unitary for any $\nu\in\mathbb R$. This is the {\it principal series representation}, $\mc P_\Delta$, and the center focus of this paper. We can fix the norm of these representations to
\beq\label{eq:PSnorm}
    \langle r|r'\rangle_{\mc P_\Delta}=\delta_{rr'}~,\qquad r\in\mathbb Z~\text{ or }~r+\frac12\in\mathbb Z~.
\eeq

For $\Delta\in\mathbb R$ we proceed by cases. If we assume that $\Delta\notin\mathbb Z$ then positivity of \eqref{eq:normratio} requires both $r\in\mathbb Z$ and $\Delta\in(0,1)$. This is the {\it complementary series representation}, $\mc C_\Delta$, and it exists with only even parity. Solving the recurrence relation\footnote{The overall coefficient is fixed by requiring \eqref{eq:CSnorm} and \eqref{eq:PSnorm} agree at the intersection of the principal and complementary series at $\Delta=\frac12$.} set by \eqref{eq:normratio} gives the norm
\beq\label{eq:CSnorm}
    \langle r|r'\rangle_{\mc C_\Delta}=\frac{\Gamma(r+1-\Delta)}{\Gamma(r+\Delta)}\delta_{rr'}~,\qquad r\in\mathbb Z~.
\eeq

Lastly, when either $\Delta\in\mathbb Z_+$ and $r\in\mathbb Z$, or $\Delta\in\mathbb N+\frac12$ and $r+\frac{1}{2}\in\mathbb Z$, then from \eqref{eq:Laction} we see there are states at $r=\mp\Delta$ which are annihilated by $l_\pm$, respectively, and upon which we terminate the representation. These are the {\it highest} and {\it lowest weight discrete series representations}, $\mc D^+_\Delta$ and $\mc D^-_\Delta$, with $r\leq -\Delta$ and $r\geq \Delta$, respectively. It is easy to check that for the ranges of $r$ allowed by the representation, \eqref{eq:normratio} is positive. These representations have norm
\beq
    \langle r|r'\rangle_{\mc D^\pm_\Delta}=\frac{\Gamma(\mp r+1-\Delta)}{\Gamma(\mp r+\Delta)}\delta_{rr'}~,\qquad r=\mp\Delta,\mp(\Delta+1),\mp(\Delta+2),\ldots
\eeq
Lastly for both the principal and complementary series, representations labelled by $\Delta$ are isomorphic to their {\it shadow representation} labelled by conformal dimension $\bar\Delta=1-\Delta$. Thus the full spectrum of principal series representations are given by $\Delta=\frac{1}{2}+i\nu$ with $\nu>0$ and the full spectrum of complementary series representations are given $\Delta\in(0,\frac12)$.

\section{Orthogonality of coordinate transformation}\label{app:coordtrafo}

As in equation \eqref{eq:relativcoorddef}, we introduce the coordinate transformation
\beq 
y_k\equiv \alpha_k\,x_{k+1}+\beta_k\sum_{j=1}^k x_j~, \qquad 1\leq k \leq N-1~, \qquad y_N = \frac{1}{\sqrt{N}}\sum_{i=1}^N x_i~,
\eeq
with coefficients 
\beq \label{eq:coefficients}
\alpha_k \equiv \sqrt{\frac{k}{k+1}}~, \qquad\qquad \beta_k \equiv - \frac{1}{\sqrt{k(k+1)}}~,
\eeq
so that the above coordinate transformation can be written in matrix form as
\beq 
y_k = \sum_{i=1}^{N} B_{ki} x_i~,
\eeq
where the matrix $B\equiv A^{-1}$ (see \eqref{eq:basischangey}) has non-zero entries given by
\beq
    B_{Ni}=\frac{1}{\sqrt N}~,\qquad B_{ki}=\begin{cases}
        \beta_k ~, \hspace{10pt} &i\leq k~, \\
        \alpha_k ~, \hspace{10pt} & i=k+1~, \hspace{20pt} (k<N)\\
        0~, \hspace{10pt} & i>k+1 ~.
    \end{cases}
\eeq
We want to show that the matrix $B$ is orthogonal, meaning:
\beq\label{eq:Borthog}
    \qquad \left(BB^T\right)_{jk}=\sum_{i=1}^NB_{ji}B_{ki}=\delta_{jk}~,\qquad \text{and} 
    \left(B^TB\right)_{jk}=\sum_{i=1}^NB_{ij}B_{ik}=\delta_{jk}~.
\eeq
Let us start with the first expression in \eqref{eq:Borthog}. Due the to the manifest symmetry of $j\leftrightarrow k$ of the above expression, it suffices to consider cases where $j\leq k$. First consider the case when $k=N$:
\beq
   \left(BB^T\right)_{jN}=\frac{1}{\sqrt{N}}\sum_{i=1}^NB_{ji}~,
\eeq
then if $j=N$ it is easy to see that $\left(BB^T\right)_{NN}=1$. However if $j<N$ then
\beq
   \left(BB^T\right)_{jN}=\frac{1}{\sqrt{N}}\left(j \beta_j+\alpha_j\right)=0~.
\eeq

Next consider the case that $k<N$. Since the entries in the matrix $B_{ki}$ with $i>k+1$ are zero we can write:
\beq \label{eq:BBT}
\left(BB^T\right)_{jk}=B_{j(k+1)}B_{k(k+1)}+\sum_{i=1}^k B_{ji}B_{ki}~.
\eeq  
If $j<k$ then $B_{j(k+1)}=0$ and the first term on the right-hand side is zero, while the second term is only non-zero when $i\leq j+1$, hence: 
\beq 
\left(BB^T\right)_{jk} =B_{j(j+1)}B_{k(j+1)}+\sum_{i=1}^{j} B_{ji}B_{ki}=\alpha_j\beta_k+j\beta_j\beta_k=0~. 
\eeq

Lastly for the diagonal entries with $j=k<N$,
\beq
    \left(BB^T\right)_{kk}=\left(B_{k(k+1)}\right)^2+\sum_{i=1}^k (B_{ki})^2=(\alpha_k)^2+k(\beta_k)^2=1~.
\eeq
Hence, we have shown that 
\beq 
\left(BB^T\right)_{jk} = \delta_{jk}~.
\eeq
A similar computation shows that $\left(B^TB\right)_{jk}=\delta_{jk}$, and we conclude that the matrix $B=A^{-1}$ is orthogonal. Its transpose $B^T$ is the matrix $A$ that is defined in the main text, which, as we have just shown, is also orthogonal.

\section{Orthogonality of associated Legendre functions}

\label{app:B}

Let us explicitly check the orthogonality relation \eqref{eq:orthoY}. The associated Legendre functions $P_{i\sigma-\frac12}^{\lambda-\frac12}(u)$ satisfy the following differential equation:
\beq\label{eq:haLPeq}
\frac{\dd}{\dd u}\left((1-u^2)\frac{\dd}{\dd u}P_{i\sigma-\frac12}^{\lambda-\frac12}(u)\right)+\left[\left(i\sigma-\frac12\right)\left(i\sigma+\frac12\right)-\frac{(\lambda-\frac12)^2}{1-u^2}\right]\haLP{\sigma}(u)=0~.
\eeq
In the following, we assume that $u>1$. We want to investigate these functions under the following inner product
\beq \label{eq:AinnerproductP}
\Big\langle \haLP{\sigma_1},\haLP{\sigma_2}\Big\rangle\equiv\int_1^\infty \dd u\,\haLP{\sigma_1}(u)\haLP{\sigma_2}(u)~,
\eeq
and show that they are orthogonal. To that end, let us multiply $\haLP{\sigma_1}(u)$ with equation \eqref{eq:haLPeq} for $\haLP{\sigma_2}(u)$ and subtract $\haLP{\sigma_2}(u)$ times equation \eqref{eq:haLPeq} for $\haLP{\sigma_1}(u)$.  After rearranging, we find that
\beq
(\sigma_1^2-\sigma_2^2)\haLP{\sigma_1}\haLP{\sigma_2}=\haLP{\sigma_2}\frac{\dd}{\dd u}\left[(1-u^2)\frac{\dd\haLP{\sigma_1}}{\dd u}\right]-\haLP{\sigma_1}\frac{\dd}{\dd u}\left[(1-u^2)\frac{\dd\haLP{\sigma_2}}{\dd u}\right]~.
\eeq
We will now integrate both sides from $u=1$ to $u=R$, and take the limit $R\rightarrow\infty$ afterwards. Integrating both terms by parts once, one finds the integrands cancel and the result is a boundary term:
\beq\label{eq:innerprod1}
\int_1^R\dd u\,\haLP{\sigma_1}(u)\haLP{\sigma_2}(u)=\frac{1-R^2}{\sigma_1^2-\sigma_2^2}\left(\haLP{\sigma_2}(R)\frac{\dd}{\dd u}\haLP{\sigma_1}(R)-\haLP{\sigma_1}(R)\frac{\dd}{\dd u}\haLP{\sigma_2}(R)\right)~.
\eeq
Here, we have dropped the boundary term at $u=1$ assuming that the functions are regular there when we take $(1-u^2)\rightarrow0$. This assumption is true when $\lambda<\frac{3}{2}$.  

We want to study \eqref{eq:innerprod1} in the $R\rightarrow\infty$ limit. Given that the associated Legendre functions are labelled by a continuous variable $\sigma$, we expect any orthogonality relation amongst them to be distributional in form, so to this end we will integrate \eqref{eq:innerprod1} against a test function. We consider 
\begin{align}
\int_{-\infty}^\infty &\dd\sigma_1\,f(\sigma_1)\Big\langle\haLP{\sigma_1},\haLP{\sigma_2}\Big\rangle \nonumber \\
&=\lim_{R\rightarrow\infty}\int_{-\infty}^{\infty}\dd\sigma_1\,f(\sigma_1)\frac{(1-R^2)}{\sigma_1^2-\sigma_2^2}\left(\haLP{\sigma_2}(R)\frac{\dd}{\dd u}\haLP{\sigma_1}(R)-\haLP{\sigma_1}(R)\frac{\dd}{\dd u}\haLP{\sigma_2}(R)\right)~. \label{eq:testfunction}
\end{align}
We will not rederive the asymptotics of the functions on the right-hand side, but use the result of \cite{van1954orthogonality}, namely that as $R\to \infty$ the associated Legendre functions satisfy
\beq 
\haLP{\sigma}(R) \sim \frac{1}{\sqrt{R}}\left( A(\sigma) \cos(\sigma\log R)+B(\sigma)\sin(\sigma \log R)\right)(1+\mathcal{O}(1/R^2))~,
\eeq
where the coefficients in the above expression are given by 
\beq \label{eq:ABcoefficients}
A(\sigma)\equiv2\text{Re}\left(\frac{2^{i\sigma-\frac12}\Gamma(i\sigma)}{\sqrt\pi\Gamma\left(i\sigma+1-\lambda\right)}\right)~, \qquad B(\sigma)\equiv -2\text{Im}\left(\frac{2^{i\sigma-\frac12}\Gamma(i\sigma)}{\sqrt\pi\Gamma\left(i\sigma+1-\lambda\right)}\right).
\eeq
Using the above asymptotics in  the $R\rightarrow\infty$ limit, the right-hand side of \eqref{eq:testfunction} can be reorganized as
\begin{align}
\lim_{R\rightarrow\infty}\int_{-\infty}^{\infty} \dd\sigma_1f(\sigma_1)&\left(\frac{\varphi_1}{\sigma_1^2-\sigma_2^2}\sin(\sigma_1\log R)\sin(\sigma_2\log R)+\frac{\varphi_2}{\sigma_1^2-\sigma_2^2}\cos(\sigma_1\log R)\cos(\sigma_2\log R)\right.\nonumber\\
&\left.+\frac{\varphi_3}{\sigma_1+\sigma_2}\sin(\sigma_1\log R)\cos(\sigma_2\log R)+\frac{\varphi_4}{\sigma_1^2-\sigma_2^2}\sin((\sigma_1-\sigma_2)\log R)\right)~,
\end{align}
where we have dropped a factor of $(R^2-1)/R^2$ in the above expression as we take $R\to \infty$, and we have introduced the following combinations:
\begin{align}
\varphi_1 & \equiv \, \sigma_1A(\sigma_1)B(\sigma_2)-\sigma_2A(\sigma_2)B(\sigma_1)~, &  \quad \varphi_2\equiv & \, \sigma_2A(\sigma_1)B(\sigma_2)-\sigma_1A(\sigma_2)B(\sigma_1)~,\nonumber\\
\varphi_3 & \equiv A(\sigma_1)A(\sigma_2)-B(\sigma_1)B(\sigma_2)~, & \quad 
\varphi_4\equiv & \, \sigma_1A(\sigma_1)A(\sigma_2)+\sigma_2B(\sigma_1)B(\sigma_2)~.
\end{align}
We now decompose the sine and cosine functions via the standard rules
\begin{align}
\sin(\sigma_1\log R)\sin(\sigma_2\log R)= & ~ \frac{1}{2}\left(\cos((\sigma_1-\sigma_2)\log R)-\cos((\sigma_1+\sigma_2)\log R)\right)~,\\
\cos(\sigma_1\log R)\cos(\sigma_2\log R)= & ~ \frac{1}{2}\left(\cos((\sigma_1+\sigma_2)\log R)+\cos((\sigma_1-\sigma_2)\log R)\right)~,\\
\sin(\sigma_1\log R)\cos(\sigma_2\log R)= & ~ \frac12\left(\sin((\sigma_1+\sigma_2)\log R)-\sin((\sigma_1-\sigma_2)\log R)\right)~. \label{eq:sincos}
\end{align}
Note that against smooth functions of $\sigma_1$, $\sin((\sigma_1\pm\sigma_2)\log R)$ and $\cos((\sigma_1\pm\sigma_2)\log R)$ integrate to zero as $R\rightarrow\infty$ due to phase cancellations. We do have to be more careful however about singular behavior as $\sigma_2\rightarrow\pm\sigma_1$ since in that case the phase vanishes. We note the following
\begin{align}
\lim_{\sigma_2\rightarrow\sigma_1}\varphi_1=~&0~,&\lim_{\sigma_2\rightarrow-\sigma_1}\varphi_1=~&0~, \\
\lim_{\sigma_2\rightarrow\sigma_1}\varphi_2= ~&0~,&\lim_{\sigma_2\rightarrow-\sigma_1}\varphi_2=~&0~, \\
\lim_{\sigma_2\rightarrow\sigma_1}\varphi_3=~ &0~,& \lim_{\sigma_2\rightarrow-\sigma_1}\varphi_3=~&A(\sigma_1)^2+B(\sigma_1)^2~, \\
\lim_{\sigma_2\rightarrow\sigma_1}\varphi_4=~&\sigma_1(A(\sigma_1)^2+B(\sigma_1)^2)~,&\lim_{\sigma_2\rightarrow-\sigma_1}\varphi_4=~&\sigma_1(A(\sigma_1)^2+B(\sigma_1)^2)~,
\end{align}
so it suffices to look at the $\varphi_3$ and $\varphi_4$ terms. We find a contribution from $\varphi_3$ as $\sigma_2\rightarrow-\sigma_1$. Using the decomposition \eqref{eq:sincos} and the distributional limit 
\beq\label{eq:deltalimit}
\lim_{a\rightarrow\infty}\frac{\sin(ax)}{x}=\pi\delta(x)~,
\eeq
one finds that
\beq 
\lim_{R\to \infty}\frac{\varphi_3}{\sigma_1+\sigma_2}\sin(\sigma_1\log R)\cos(\sigma_2\log R)=\frac{\pi}{2}\left(A(\sigma_1)^2+B(\sigma_1)^2\right)\delta(\sigma_1+\sigma_2)~.
\eeq
Using the definition of the coefficients $A(\sigma)$ and $B(\sigma)$ in \eqref{eq:ABcoefficients} the factor multiplying the delta-function can easily computed to be 
\beq
\frac{\pi}{2}\left(A(\sigma_1)^2+B(\sigma_1)^2\right)=\frac{\Gamma(i\sigma_1)\Gamma(-i\sigma_1)}{\Gamma(i\sigma_1+1-\lambda)\Gamma(-i\sigma_1+1-\lambda)}~.
\eeq
Similarly, one has to include the contribution from $\varphi_4$ as $\sigma_2\rightarrow\sigma_1$. We have 
\beq 
\lim_{R\to \infty}\frac{\varphi_4}{\sigma_1+\sigma_2}\frac{\sin((\sigma_1-\sigma_2)\log R)}{\sigma_1-\sigma_2} = \frac{\pi}{2}(A(\sigma_1)^2+B(\sigma_1)^2)\delta(\sigma_1-\sigma_2)~.
\eeq
We then arrive at the following orthogonality relation\footnote{Note that there is a slight discrepancy with the result in \cite{van1954orthogonality}, since the term involving the delta-function $\delta(\sigma_1+\sigma_2)$ does not appear there and additionally, the case they studied was $\lambda-\frac{1}{2}\in\mathbb{Z}$.}
\beq \label{eq:orthogonalityP}
\Big\langle \haLP{\sigma_1},\haLP{\sigma_2}\Big\rangle=\frac{\Gamma(i\sigma_1)\Gamma(-i\sigma_1)}{\Gamma(i\sigma_1+1-\lambda)\Gamma(-i\sigma_1+1-\lambda)}\left(\delta(\sigma_1+\sigma_2)+\delta(\sigma_1-\sigma_2)\right)~.
\eeq 

\section{Orthogonality of interacting principal series wavefunctions}\label{app:PSWFsortho}

In this appendix we will demonstrate that the hyperbolic wavefunctions, $P_\gamma(u,\vec\nu)$, of the $N$-particle principal series Calogero model are plane-wave orthonormal with respect to the inner product
\beq
    (P_{\gamma_1},P_{\gamma_2})=\int_0^\infty \dd u\int_{S^{N-3}} d\Omega_{N-3}\sqrt{G}|\widehat{\vdm}|^{2\lambda}P_{\gamma_1}^\ast(u,\vec\nu)P_{\gamma_2}(u,\vec\nu)=\delta(\gamma_1-\gamma_2)~,
\eeq
up to suitable normalization. Given the shadow isomorphism on principal series representations we take both $\gamma_{1,2}$ to be positive. We will follow the basic method established in \cref{app:B}. That is, let us consider multiplying the inner product by $(\gamma_2^2-\gamma_1^2)$. Using the defining equation \eqref{eq:NPSCasEq2} this is equivalent to 
\begin{align}
   \big (\gamma_2^2-&\gamma_1^2\big)(P_{\gamma_1},P_{\gamma_2})=\nonumber\\
    &\int_0^\infty \dd u\int_{S^{N-3}}\dd\Omega_{N-3}\left[\pa_\mu\left(\sqrt{G}G^{\mu\nu}|\widehat{\vdm}|^{2\lambda}\pa_\nu P_{\gamma_1}\right)^\ast P_{\gamma_2}-P_{\gamma_1}^\ast \pa_\mu\left(\sqrt{G}G^{\mu\nu}|\widehat{\vdm}|^{2\lambda}\pa_\nu P_{\gamma_2}\right)\right]~.
\end{align}
Integrating each of the terms on the right-hand side once, it is easy to see that for $\gamma_1\neq\gamma_2$ the resulting integrands cancel and
\beq
    (P_{\gamma_1},P_{\gamma_2})=0~,\qquad \gamma_1\neq\gamma_2~.
\eeq
For coincident $\gamma$'s we will need to be more careful with the boundary terms. For $N>3$ both $\sqrt{G}$ and $\widehat{\vdm}$ vanish at $u=0$; we will put in an artificial boundary at $u=\Lambda$ and take the $\Lambda\rightarrow\infty$ limit:
\begin{align}
    (\gamma_2^2-&\gamma_1^2)(P_{\gamma_1},P_{\gamma_2})=\lim_{\Lambda\rightarrow\infty}\int_{S^{N-3}}\dd\Omega_{N-3}\left[\sqrt{G}\,|\widehat{\vdm}|^{2\lambda}\left(\left(\pa_u P_{\gamma_1}\right)^\ast P_{\gamma_2}-P_{\gamma_1}^\ast\,\pa_u P_{\gamma_2}\right)\right]_{u=\Lambda}~.
\end{align}
Asympotically $\widehat{\vdm}$ takes the form
\beq
    \widehat{\vdm}\sim \left(\frac{e^u}{2}\right)^{\frac{N(N-1)}{2}}\prod_{a}\mathbf{b}^a\cdot \hat{\tilde{\mathbf{z}}}\equiv\left(\frac{e^u}{2}\right)^{\frac{N(N-1)}{2}}\widehat{\delta}~,\qquad \hat{\tilde{\mathbf{z}}}=(1,i\vec\nu)~,
\eeq
where `$\sim$' means `equal as $u\rightarrow\infty$.' Note that $\hat{\delta}$ is a function only of the $S^{N-3}$ coordinates and either purely real or purely imaginary:
\beq
    \hat\delta=(-1)^{N-2}i^{\frac{(N-3)(N-2)}{2}}\left[\prod_{j>2}\left(\frac{1}{4}+\zeta^{[j,1]2}\right)\right]\left[\prod_{i>j>2}\zeta^{[i,j]}\right]~.
\eeq
We will need the asymptotic forms of the hyperbolic wavefunctions $P_{\gamma}$. At large $u$, the defining equation for $P_{\gamma}$ takes the form
\beq\label{eq:limitNPSang}
    \left[\pa_u^2+2\lambda_N\pa_u+\frac{e^{-2u}}{4}\frac{\hat{\delta}^{-2\lambda}}{\sqrt{g_{S^{N-3}}}}\pa_i\left(\hat{\delta}^{2\lambda}\sqrt{g_{S^{N-3}}}g^{ij}_{S^{N-3}}\pa_j\right)+\left(\lambda_N^2+\gamma^2\right)\right]P_\gamma\sim0~,
\eeq
where for convenience we have defined
\beq
    \lambda_N\equiv \lambda\frac{N(N-1)}{2}+\frac{N-3}{2}~.
\eeq
The sphere portion of this equation is a real linear deformation of the $S^{N-3}$ Laplacian
\beq
    \frac{\hat{\delta}^{-2\lambda}}{\sqrt{g_{S^{N-3}}}}\pa_i\left(\hat{\delta}^{2\lambda}\sqrt{g_{S^{N-3}}}g^{ij}_{S^{N-3}}\pa_j\right)=\nabla_{S^{N-3}}^2+\lambda\hat{\mc M}~,\qquad \hat{\mc M}=2g^{ij}_{S^{N-3}}\pa_i\log\hat\delta\,\pa_j~,
\eeq
and is reminiscent of the angular operator \eqref{eq:Mhat} appearing in the relative angular dynamics of a $(N-1)$-particle ordinary Calogero model. However this operator is not the same as $\hat{\delta}$ involves the positive roots of $A_{N-1}$ as opposed to $A_{N-2}$. Regardless, this operator is Hermitian with respect to the inner product
\beq
    (\Phi_1,\Phi_2)_{S^{N-3}}=\int_{S^{N-3}}\dd\Omega_{N-3}\sqrt{g_{S^{N-3}}}|\hat{\delta}|^{2\lambda}\,\Phi_1(\vec\nu)^\ast\Phi_2(\vec\nu)~,
\eeq
and we will suppose that it admits an orthonormal set of discrete eigenfunctions labelled by a multi-index $\vec\ell$,\footnote{When $\lambda$=0, $\vec\ell=(\ell,\vec m)$, the quantum numbers of the hyperspherical harmonics.} $\hat{\mathcal Y}_{\vec\ell}$, with eigenvalue
\beq
    \frac{\hat{\delta}^{-2\lambda}}{\sqrt{g_{S^{N-3}}}}\pa_i\left(\hat{\delta}^{2\lambda}\sqrt{g_{S^{N-3}}}g^{ij}_{S^{N-3}}\pa_j\hat{\mathcal Y}_{\vec\ell}\right)=-\varepsilon_{\vec\ell}\hat{\mc Y}_{\vec\ell}~,
\eeq
and normalized to
\beq
    (\hat{\mc Y}_{\vec\ell_1},\hat{\mc Y}_{\vec\ell_2})_{S^{N-3}}=\delta_{\vec\ell_1,\vec\ell_2}~.
\eeq
We will write asymptotically $P_\gamma(u,\vec\nu)\sim{\mathsf c}_{\vec\ell}\,\rho_\gamma^{\vec\ell}(u)\hat{\mc Y}_{\vec\ell}(\vec\nu)$ with $\sum_{\vec\ell}|\mathsf c_{\vec\ell}|^2=1$. In the large $u$ limit however, we do not need to know the $S^{N-3}$ eigenvalue, $\varepsilon_{\vec\ell}$, as it is exponentially suppressed in \eqref{eq:limitNPSang}. In this limit, $\rho_\gamma^{\vec\ell}(u)$ satisfies
\beq
    \left[\pa_u^2+2\lambda_N\pa_u+\lambda_N^2+\gamma^2\right]\rho_\gamma^{\vec\ell}(u)\sim0~,
\eeq
for all $\vec\ell$ and admits decaying plane-wave solutions:
\beq
    \rho_\gamma^{\vec\ell}(u)\sim \left(\frac{e^{u}}{2}\right)^{-\lambda_N}\,\left(\mathsf{a}_{\vec\ell,\gamma}^{(+)}\,e^{i\gamma u}+\mathsf{a}^{(-)}_{\vec\ell,\gamma}\,e^{-i\gamma u}\right)~,
\eeq
for arbitrary coefficients $\mathsf{a}^{(\pm)}_{\vec\ell,\gamma}$ and the $2^{\lambda_N}$ is a normalization for later convenience.

With this we can express the inner product on the hyperbolic wavefunctions as
\begin{align}
    (P_{\gamma_1},P_{\gamma_2})=&\frac{1}{(\gamma_2^2-\gamma_1^2)}\lim_{\Lambda\rightarrow\infty}\left(\frac{e^\Lambda}{2}\right)^{2\lambda_N}\sum_{\vec\ell}|{\mathsf c}_{\vec\ell}|^2\left[(\pa_u\rho^{\vec\ell}_{\gamma_1})^\ast\rho_{\gamma_2}^{\vec\ell}-(\rho_{\gamma_1}^{\vec\ell})^{\ast}\pa_u\rho_{\gamma_2}^{\vec\ell}\right]_{u=\Lambda}\nonumber\\
    =&\lim_{\Lambda\rightarrow\infty}\sum_{\vec\ell}|{\mathsf c}_{\vec\ell}|^2\left[\mathsf{a}_{\vec\ell,\gamma_1}^{(+)\ast}\mathsf{a}_{\vec\ell,\gamma_2}^{(+)}\frac{ie^{-i(\gamma_1-\gamma_2)\Lambda}}{(\gamma_1-\gamma_2)}-\mathsf{a}_{\vec\ell,\gamma_1}^{(-)\ast}\mathsf{a}_{\vec\ell,\gamma_2}^{(-)}\frac{ie^{i(\gamma_1-\gamma_2)\Lambda}}{(\gamma_1-\gamma_2)}\right.\nonumber\\
    &\qquad\qquad\qquad\qquad\qquad\left.+\mathsf{a}_{\vec\ell,\gamma_1}^{(+)\ast}\mathsf{a}_{\vec\ell,\gamma_2}^{(-)}\frac{ie^{-i(\gamma_1+\gamma_2)\Lambda}}{(\gamma_1+\gamma_2)}-\mathsf{a}_{\vec\ell,\gamma_1}^{(-)\ast}\mathsf{a}_{\vec\ell,\gamma_2}^{(+)}\frac{ie^{i(\gamma_1+\gamma_2)\Lambda}}{(\gamma_1+\gamma_2)}\right]~.
\end{align}
At this point we can use the limit \eqref{eq:deltalimit} to find
\beq
    (P_{\gamma_1},P_{\gamma_2})=\pi\sum_{\vec\ell}|\mathsf{c}_{\vec\ell}|^2\left(|\mathsf{a}^{(+)}_{\vec\ell,\gamma_1}|^2+|\mathsf{a}^{(-)}_{\vec\ell,\gamma_1}|^2\right)\delta(\gamma_1-\gamma_2)~,
\eeq
where we have dropped the terms proportional to $\delta(\gamma_1+\gamma_2)$ as they cannot be satisfied with the supposition of $\gamma_{1,2}$ both positive.

\bibliographystyle{utphys} 
\bibliography{calrefs.bib}{}

\end{spacing}
\end{document}